\documentclass[twocolumn]{aastex631}
\usepackage{gensymb}
\usepackage{textcomp}
\usepackage{amsmath}
\usepackage{tabularx}

\newcommand{\ha}{H$\alpha$}
\newcommand{\caII}{Ca\,{\sc ii}}
\newcommand{\chisq}{$\chi_{\nu}^2$}

\begin{document}

\title[Spectral Synthesis for CAe Stars]{Spectral Synthesis for Classical Ae Stars: \ha\ and \caII}

\correspondingauthor{R.\ Anusha, T.\ A.\ A.\ Sigut}
\email{araviku4@uwo.ca; asigut@uwo.ca}

\author[0000-0002-9015-6417]{R.\ Anusha}
\affiliation{Department of Physics \& Astronomy, The University of Western Ontario \\ 
1151 Richmond Street, London, ON, Canada, N6A~3k7}

\author[0000-0002-0803-8615]{T.\ A.\ A.\ Sigut}
\affiliation{Department of Physics \& Astronomy, The University of Western Ontario \\ 
1151 Richmond Street, London, ON, Canada, N6A~3k7}
\affiliation{Institute for Earth \& Space Exploration (IESX), The University of Western Ontario \\  7134 Perth Drive, London, ON, Canada, N6A~5B7}

\begin{abstract}
Classical Ae stars (CAe) are main sequence, A-type stars with \ha\ emission but no signature of dust; they are thought to be the cool extension of the classical Be (CBe) stars to lower stellar masses. We investigate the use of \ha\ and the \caII\ IR triplet observed at low spectroscopic resolution (${\cal R}=1800$) as diagnostics for the circumstellar disks of the CAe stars. Using a sample of 159 CAe stars from the LAMOST survey (DR5), we find that circumstellar disk masses and star/disk viewing inclination angles can be reliably determined from low resolution \ha\ profiles. The CAe disk mass distribution agrees well with the previous range found for late-type CBe stars, although there is an apparent increase in disk mass for the latest CAe spectral sub-types (A3 and A4) which we attribute to selection effects. Over the spectral range A2 through B6, the average disk-to-star mass ratio for the CAe and CBe stars is consistent with $\log_{10}(M_{\rm disk}/M_*)=-9.1$. A minority of the CAe sample ($\sim$16\%) also show emission in the Ca\,{\sc ii} IR triplet. The observed emission is well within the range predicted by the disk models, and there is statistical evidence that the \caII\ emitters have systematically larger \ha\ disk masses. 
\end{abstract}


\section{Introduction}

The classical Ae (CAe) stars are thought to represent the continuation of the classical Be (CBe) star phenomena to the cooler A spectral class \citep{Jaschek1988,Jaschek1991,Bohlender2016}. Similar to their hotter and more numerous CBe star counterparts, the CAe stars are main-sequence stars producing Balmer-line emission in a low-density, gaseous disk that forms at least once during their lifetime \citep[e.g.,][]{Andrillat1986, Jaschek1998}. Unlike Herbig Ae (HAe) stars, which are pre-main-sequence objects surrounded by dusty {\it accretion\/} disks \citep{Herbig1960, Waters1998, Brittain2023}, CAe stars host gaseous {\it decretion\/} disks. They offer a way to explore if the same physical mechanisms governing disk formation and angular momentum transport in CBe star disks also operate in the cooler A-type stars \citep{Carciofi2011a, Rivinius2013}.

The thermal structure of CBe star disks are generally understood to be controlled by photoionization heating from the central star, with radiative-equilibrium models successfully reproducing the observed disk temperatures and emission properties \citep{Carciofi2006, Sigut2007, Silaj2010a,Sigut2023}. In contrast, the lower luminosities and weaker ionizing radiation fields of CAe stars raise the possibility that radiative heating alone may not fully determine the disk thermal structure. Recent models by \cite{Anusha2025} indicate that for CAe stars of spectral type A2 and later, viscous shear heating can raise disk temperatures and enhance \ha\ emission. These results imply that viscous dissipation is an important component of the energy budget in some CAe disks and suggests that the sensitivity of \ha\ emission to viscous heating may provide an observational means for constraining the disk viscosity $\alpha$ parameter \citep{Shakura1973a}. These findings underscore the importance of obtaining robust observational constraints on CAe disks.

Despite the growing interest in these objects, CAe disks remain challenging to detect and characterize systematically, even in an era of large-scale spectroscopic surveys. This is for several, interconnected reasons: first, their spectroscopic signatures are intrinsically weak. 
\cite{Anusha2021} identified CAe stars using the Large Sky Area Multi-Object Fiber Spectroscopic Telescope (LAMOST; \citealt{Cui2012, Bai2021}) Data Release 5 (DR5), and the measured \ha\ equivalent widths range from approximately $-0.2$ to $-23.6\;$\AA\footnote{In this work, we adopt the convention that positive equivalent widths indicate net emission, and negative equivalent widths, net absorption.}, with the median of $-1.5$ {\AA}. 
At the survey's resolution ($R \approx 1800$), faint line fluxes often approach the noise floor, reducing the effective signal-to-noise ratio for disk diagnostics and increasing false negatives due to continuum-subtraction residuals and sky contamination \citep{Shridharan2021}. Secondly, the selection criteria for CAe candidates used in these surveys preferentially identify stars with prominent \ha\ emission peaks, thereby biasing the sample toward stronger emitters while under-representing objects with weaker or more complex line morphologies, such as shell-type profiles. Thirdly, there is a paucity of high-resolution spectroscopy and multi-wavelength photometry for most CAe stars. Consequently, key physical parameters of their disks such as density, mass, and outer disk radius remain poorly constrained compared to the extensive literature on CBe stars \citep{Carciofi2009, Silaj2010a, Haubois2012, Klement2017, Rimulo2018, Sigut2023}. 

This study aims to provide a uniform characterization of CAe disk properties through radiative-transfer modeling of spectra from the LAMOST CAe sample of \citet{Anusha2021}. This sample consists of 159 CAe stars (spectral types A0 through A4), all with available low-resolution (${\mathcal R}=1800$) spectra in the wavelength range $3700-9000\,$\AA. The LAMOST DR5 CAe survey increased the number of confirmed CAe stars by nearly a factor of nine, and therefore provides a large and homogeneous spectral catalog for analysis. Using the \texttt{Bedisk/Beray} spectral-synthesis code suite \citep{Sigut2018}, radiative-transfer modeling of \ha\ emission profiles will be used to estimate disk densities and masses in order to examine potential correlations with spectral subtype and to assess whether CAe stars host systematically less dense, less massive disks as compared to CBe star hosts \citep{Grundstrom2006, Touhami2013, Arcos2017}.

In addition to \ha\ emission, \citet{Anusha2021} identified emission features in the hydrogen Paschen series, Fe\,{\sc ii}, O\,{\sc i}, and the \caII\ infrared triplet (IRT). Among these, \caII\ IRT emission was the most prevalent, detected in 25 of the 159 stars ($\sim16$\%). We therefore also investigate \caII\ IRT emission as a complementary diagnostic for probing the circumstellar disks of CAe stars. The presence of \caII\ IRT emission in CBe stars is associated with other circumstellar features, including the Paschen lines and infrared excess, suggesting its origin in the dense, cooler regions of the circumstellar disk \citep{Briot1981, Andrillat1990}. Because the \caII\ IRT lines likely form in relatively dense, partially ionized gas in CBe stars, whereas \ha\ emission can form over a broader range of densities and typically traces the more extended disk, the two diagnostics are sensitive to different physical conditions and potentially different disk locations \citep{Banerjee2021}. 

A limitation to focusing on the CAe sample of \citet{Anusha2021} is that currently less is known about the broader stellar properties of the sample stars, particularly in terms parameters that might correlate with circumstellar disk formation, such as binarity, non-radial pulsation, rotation rate, and potential magnetic fields. A cross-match of the sample with the SIMBAD database \citep{Wenger2000} identified one spectroscopic binary, four eclipsing binaries, two $\delta$ Scuti variables, and 60 emission-line stars without spectra types. Only seven stars are identified as probable members of open clusters, with no reliable cluster membership information available for the remaining targets.

More is known about the relatively small number of CAe stars identified prior to \citet{Anusha2021}.
\citet{Abt1973} detected shell features in 8 of the 35 rapidly rotating A dwarfs, with mean rotation rates near $180\,\mathrm{km\,s^{-1}}$ \citep{Slettebak1982, Jaschek1988, Ohanesyan1997}. Spectral variability is commonly observed in the Balmer features of CAe/A-shell stars, with timescales in the order of years \citep{Jaschek1988, Abt2008, Iliev2013}. Binarity or multiplicity was reported for more than 60\% of the CAe/A-shell stars in \citet{Jaschek1998}. The two CAe stars discovered by \citet{Monin2003} as part of a magnetic survey of bright main sequence stars, $\kappa\;$UMa and $\nu\;$Cyg, are binaries with $v\sin i=219$ and $241\,\mathrm{km\,s^{-1}}$ respectively, although no magnetic fields were detected in these stars. Previous searches for magnetic field in CAe/A-shell stars have likewise yielded no convincing detections, with circular-polarization measurements placing upper limits of order a few hundred gauss on the longitudinal fields \citep{Clayton1980, Shorlin2002}. While ZTF (Zwicky Transient Facility) and TMTS (Tsinghua University-Ma Huateng Telescopes for Survey) observations have established two CAe stars as potential $\delta\,$Scuti stars \cite[][respectively]{Chen2020dS,Guo2024dS}, there are currently no systematic studies on the link between the CAe phenomena and pulsation (unlike for the CBe stars \citealt{Balona2020, Labadie-Bartz2022}).

The paper is organized as follows: Section~\ref{sec:Calc} describes the computation of the line profile libraries. Section~\ref{sec:results} describes the \ha\ fitting method and applies it to the CAe sample. Section~\ref{sec:dmass} discusses the disk masses for the CAe sample and compares the results to the CBe stars. Section~\ref{sec:inclination} tests the hypothesis of random viewing inclinations for the CAe sample stars, and Section~\ref{sec:ca-ii} presents the analysis of the Ca\,{\sc ii} infrared triplet emission lines. Finally, Section~\ref{sec:concl} summarizes the main findings and conclusions.

\section{\texorpdfstring{Synthetic \ha\ CA\lowercase{e} Libraries}{Synthetic CAe Libraries}}
\label{sec:Calc}

A self-consistent model of a CAe star and its circumstellar disk requires two fundamental physical inputs: (i) the radial ($R$) and vertical ($Z$) density structure of the circumstellar gas, $\rho(R,Z)$, and (ii) the radiation field of the central star, which provides the photoionizing energy that determines the disk's thermal structure, $T(R,Z)$. The stellar photoionizing radiation field adopted in this work is derived from \cite{Kurucz1993} stellar atmosphere models, with limb-darkening effects incorporated into the surface radiation field. Using these inputs, we compute grids of circumstellar disk models with the \texttt{Bedisk/Beray} code suite \citep{Sigut2007,Sigut2018}, adopting central-star $(T_{\rm eff},\log g)$ parameters representative of main-sequence A-type stars.

The \texttt{Bedisk/Beray} suite offers a self-consistent framework for modeling the circumstellar environments of CAe stars and predicting their observational signatures. \texttt{Bedisk} computes the thermal structure of a gaseous, axisymmetric disk surrounding a central star, while \texttt{Beray} uses this output to solve the radiative transfer problem to generate synthetic \ha\ (and \caII) line profiles. Together, these codes link disk physical parameters to measurable observables such as line strength, shape, and width forming the basis for deriving inclination angles, disk masses, and other structural properties.

The adopted stellar parameters for the central main-sequence, A-type stars are summarized in Table~\ref{tab:Astars}. Masses and effective temperatures ($T_{\rm eff}$) were taken from the calibrations of \citet{Gray2022}. Stellar radii were derived from the assumed surface gravities, ranging from $\log g = 3.60$ to $\log g=4.40$. A wide range of surface gravities were considered since the atmospheric pressure controls the (absorption) width of the photospheric \ha\ profile for the A-type stars. Luminosities were calculated from the stellar $T_{\rm eff}$ and radii. Critical rotational velocities, computed from the stellar masses and radii following \citet{Maeder2009a}, are also given in Table~\ref{tab:Astars}. 

\begin{table}[tb]
\caption{Stellar Parameters for Main Sequence A Stars}
\label{tab:Astars}
\begin{center}
\begin{tabular}{lccccrc}
\hline \hline
SpT$^a$ &  $T_{\text{eff}}\;^b$ & $\log(g)^{\;c}$ & $R_*$ & $M_*\;^{b}$ & $L_*$ & $v_{\rm crit}$ \\
  &  (K) & & $(R_\odot)$ & ($M_\odot$) & ($L_\odot$) & ($\rm km\,s^{-1}$) \\
\hline
A0  & 9600 & 4.40 & 1.64 & 2.46 & 20.5 & 437 \\
A0  & 9600 & 4.00 & 2.60 & 2.46 & 51.4 & 347 \\
A0  & 9600 & 3.60 & 4.12 & 2.46 & 129.0  & 276 \\
A1  & 9200 & 4.40 & 1.59 & 2.31 & 16.2 & 430 \\
A1  & 9200 & 4.00 & 2.52 & 2.31 & 40.7 & 341 \\
A1  & 9200 & 3.60 & 3.99 & 2.31 & 102.0  & 271 \\
A2  & 9000 & 4.40 & 1.55 & 2.21 & 14.2 & 426  \\
A2  & 9000 & 4.00 & 2.46 & 2.21 & 35.7 & 338  \\
A2  & 9000 & 3.60 & 3.90 & 2.21 & 89.6 & 268  \\
A3  & 8600 & 4.40 & 1.53 & 2.15 & 11.5 & 422  \\
A3  & 8600 & 4.00 & 2.43 & 2.15 & 28.9 & 335  \\
A3  & 8600 & 3.60 & 3.85 & 2.15 & 45.8 & 267  \\
A4  & 8400 & 4.40 & 1.51 & 2.10 & 10.2 & 420  \\
A4  & 8400 & 4.00 & 2.40 & 2.10 & 25.7 & 334  \\
A4  & 8400 & 3.60 & 3.80 & 2.10 & 64.6 & 265  \\ \hline     
\end{tabular}
\end{center}
{\it Notes:} $(a)$ Spectral type. (b) Adopted from Appendix~B of \cite{Gray2022}. $(c)$ Models for $\log(g)=3.8$ and $4.2$ were also used but are not included in this table.
\end{table}

Each stellar model in Table~\ref{tab:Astars} is surrounded by an axisymmetric disk with the (mass) density
\begin{equation}
    \rho(R,Z) = \rho_0 \left(\frac{R}{R_*}\right)^{-n}\,\exp\left(-\left[\frac{Z}{H(R)} \right]^2\right)\;.
\label{eq:rho}
\end{equation}
Here $\rho_0$ is the equatorial density at the disk's inner radius, $R$ is the radial distance from the star ranging between $R_* \leq R \leq R_{\rm disk}$, $Z$ is the vertical coordinate (parallel to the star's rotation axis), and $n$ is the radial density exponent. The scale height $H(R)$ is fixed by assuming the disk is in vertical (Z) hydrostatic equilibrium, which leads to the condition
\begin{equation}
\frac{H(R)}{R} = \frac{c_s}{V_{K}}R\; 
\label{eq:scale-height1}
\end{equation}
where $c_s$ is the local sound speed and $V_{K}$ is the Keplerian orbital velocity. If the disk is isothermal, say with $T_{\rm disk}=0.6\,T_{\rm eff}$, then the scale height becomes
\begin{equation}
H(R) =H_0 \left(\frac{R}{R_*} \right)^{3/2}\;.
\label{eq:scale-height2}
\end{equation}
Here $H_0$ is the disk scale height at the inner edge of the disk, and the disk flares as $H\,\propto R^{3/2}$. While the disks in this work are non-isothermal, with $T(R,Z)$ determined via radiative equilibrium (see below), we have adopted Equation~(\ref{eq:scale-height2}) to fix the density structure $\rho(R,Z)$ of the disks. A more careful treatment that integrates the hydrostatic equilibrium equation along with disk temperatures \citep{Sigut2009} shows that the effects on computed \ha\ profiles are small. Finally, the assumed Keplerian rotation of the disk, $V_K = (GM_*/R)^{1/2}$, defines the disk’s velocity field and controls the Doppler broadening of the computed line profiles. 

The total disk mass associated with each model follows directly from $\rho(R,Z)$ via
\begin{equation}
    \label{eq:diskmass}
    M_{\rm disk} = 2\pi\,\int_{R_*}^{R_{\rm disk}} \int_{-\infty}^{+\infty}\; \rho(R,Z)\,R\,dZ\,dR \;,
\end{equation}
and is fixed by the parameters $(\rho_0,n,R_{\rm disk})$.

Given the disk density structure $\rho(R,Z)$, \texttt{Bedisk} calculates the temperature distribution $T(R,Z)$ by solving the coupled equations of statistical and radiative equilibrium, assuming a solar chemical composition, a standard choice in CBe disk studies \citep[e.g.,][]{Carciofi2006}. This includes photoionization and recombination of hydrogen, heating from stellar irradiation, and cooling via line emission. We have also included shear heating in the disk assuming a viscosity parameter \citep{Shakura1973a} of $\alpha=0.1$ following \citet{Anusha2025}. This level of shear heating has only a small effect on the temperature structure of disks around A0/A1 stars, but it does prevent the disks around the later A3/A4 stars from becoming very cool (see \citet{Anusha2025} for further details).

Using this density and temperature structure, the \texttt{Beray} code \citep{Sigut2018} computes the emergent \ha\ profile by integrating the radiative transfer equation along rays directed at the observer. These rays fix the viewing inclination angle $i(^\circ)$ for the observer, from 0$^{\circ}$ (pole-on star / face-on disk) to 90$^{\circ}$ (equator-on star / edge-on disk). The viewing inclination strongly affects the morphology of the \ha\ emission line profile due to the presence or absence of disk material along the line of sight which, when combined with Doppler shifts due to disk rotation, can modify both line shape and peak separation. Observationally, systems with $i \gtrsim 80^\circ$ often exhibit deep central absorption in \ha; if the emission component is weak, such stars will be identified as A-shell stars. For rays terminating on the stellar surface, an appropriately Doppler-shifted, LTE photospheric \ha\ absorption profile, corresponding to the star’s $T_{\rm eff}$ and $\log g$, is used as the upwind boundary condition. This approach naturally produces synthetic \ha\ profiles for the star+disk system which reduce to the photospheric profile in the limit $\rho_0 \to 0$.

The full grid and range of model parameters can be found in Table~\ref{tab:model_params}. Each $(T_{\rm eff},\log g)$ combination of Table~\ref{tab:Astars} is represented by 18,810 individual \ha\ line profiles, all for the possible combinations of disk density parameters $(\rho_0,n,R_{\rm disk})$ and viewing inclinations $(i)$. Each \ha\ profile was convolved down to a resolution of ${\cal R}=1800$ (or $167\,\rm km\,s^{-1})$ for comparison to the LAMOST spectra.

\begin{table}[tb]
\caption{Range and Number of Model Parameters}
\label{tab:model_params}
\begin{center}
\begin{tabular}{lll} \hline\hline
\noindent
Parameter & Range & Number, Step \\ \hline
$\rho_0\,(\rm{g\,cm}^{-3})$ & $10^{-12}-2.5\times 10^{-10}$ & 15, $\Delta\log\rho_0=0.17$ \\ 
$n$ &  $1.50-4.5$ & 11, $\Delta n=0.25$ \\
$R_{\rm disk}(R_*)$ & 5, 15, 25, 50, 65, 80 & 6, $-$  \\
$ i(^\circ)$ & $0-90^\circ$ & 19, $\Delta i=5^\circ$ \\
$\log(g)$ & $3.60-4.40$ & 5, $\Delta\log g=0.20$ \\ \hline
\end{tabular}
\end{center}
\end{table}

\section{H\texorpdfstring{$\alpha$}{alpha} Line Profile Fits}
\label{sec:results}

Among the hydrogen Balmer lines, \ha\ is particularly valuable because of its sensitivity to disk density, geometry, and ionization structure \citep{Rivinius2013,Vieira2017,Anusha2025}. By comparing observed \ha\ profiles to synthetic spectra generated with the \texttt{Bedisk/Beray}, we can assess the extent to which an axisymmetric, Keplerian decretion disk model accounts for the diversity of CAe \ha\ morphologies. Successful fits validate the model’s applicability for extracting key parameters such as the disk mass, density structure and the system viewing inclination. Mismatches may highlight where additional physics or non-axisymmetry may be required. The \ha\ fitting method here is very similar to that employed by \citet{Sigut2023} for a similarly-sized sample of CBe stars.

\begin{figure}
\centering
\includegraphics[width=1.0\columnwidth]{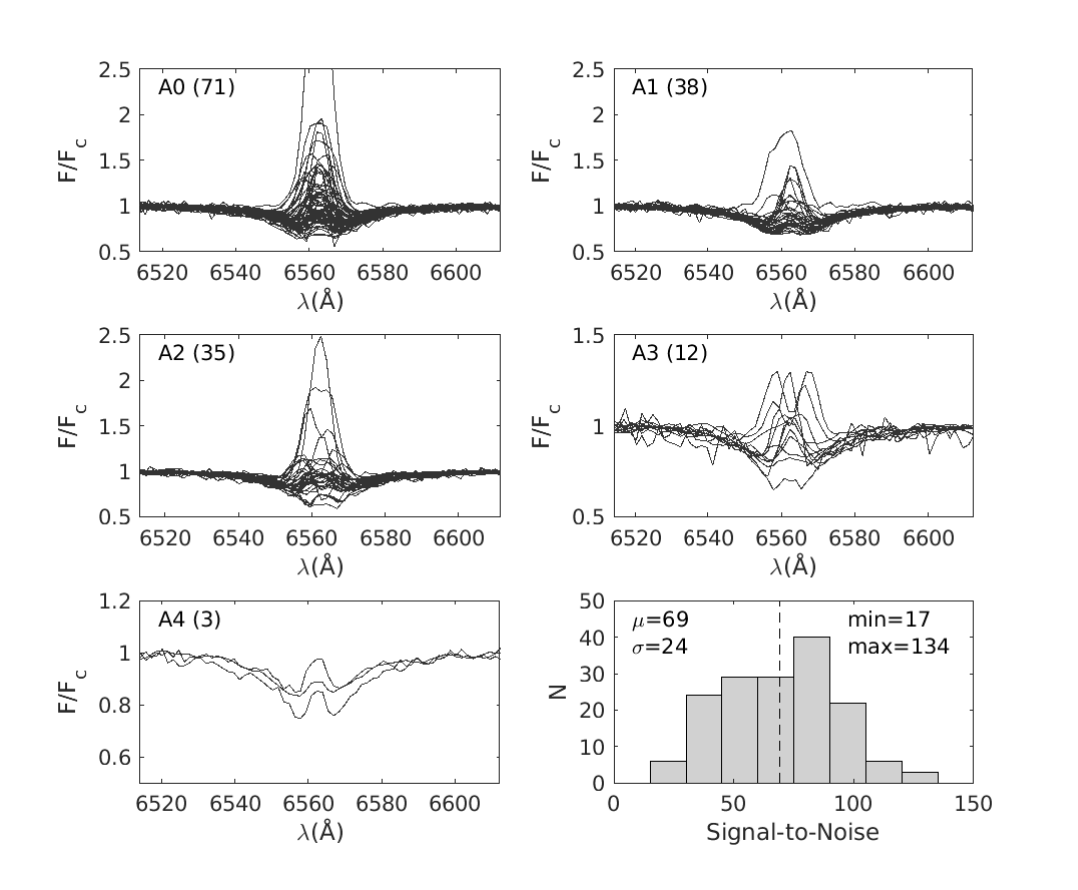}
\caption{The LAMOST (DR5) sample of 159 low-resolution (${\cal R}=1800$) CAe spectra, organized by spectral type. The number of spectra shown in each panel is indicated in brackets after the spectral type. The bottom right panel is a histogram of the sample signal-to-noise ratios and is consistent with a Gaussian of $\mu=69$ and $\sigma=24$.}
\label{fig:Obs_Halpha}
\end{figure}

Figure~\ref{fig:Obs_Halpha} shows the 159 observed CAe \ha\ profiles that form the sample, all taken from low-resolution $({\cal R}=1800)$ LAMOST (DR5) spectra \citep{Anusha2021}. A list of the sample stars with LAMOST and Gaia DR3 identifications is given in the Appendix. In Figure~\ref{fig:Obs_Halpha}, the observations are categorized into individual spectral types, ranging from A0 through A4, with the majority of sample stars (69\%) having the earliest A0 or A1 spectral type. The bottom-right panel shows the signal-to-noise ratio (S/N) distribution of the sample, estimated from the continuum adjacent to \ha, which ranges from a minimum of 17 through a maximum of 134. A Kolmogorov-Smirnov (KS) test \citep{Wall2003} indicates that this (S/N) distribution is consistent with a normal distribution of mean $\mu=69$ and standard deviation $\sigma=24$.

The maximum relative \ha\ emission strength decreases toward later spectral types (peak fluxes $\sim 2.5$ times the continuum in the A0 group versus $\sim 1.2$ in A3). The overall \ha\ profile morphologies are as follows: 77\% are single-peaked, 16\% are symmetric and double-peaked, and 7\% are asymmetric and double-peaked. The observations further reflect the generally weaker and narrower emission cores that characterize CAe disks relative to the CBe stars \citep{Anusha2021}.

To fit each \ha\ profile, the library of 94,050 profiles representing the various combinations the model parameters and $\log(g)$ values given in Table~\ref{tab:model_params} for the star's spectral-type was used. We found that allowing $\log(g)$ to vary over the main sequence range resulted in better representation of the photospheric \ha\ line wings and improved the quality of the fits. If ${\cal F}_{\lambda}^{\rm obs}$ is the continuum normalized, observed LAMOST profile (within $\pm 1250\,\rm km\,s^{-1}$ of line centre) and $F_{\lambda}^{\vec{x}}$ is the model flux for the combination of parameters represented by 
$\vec{x}\equiv(\rho_0,n,R_{\rm disk},i;\log g)$, these two profiles are compared via the reduced chi-squared statistic defined as
\begin{equation}
\chi_\nu^2(\vec{x})\equiv \frac{1}{\nu}\,\sum_{j=1}^{N_{\lambda}}\,\left(\frac{{\cal F}_{\lambda_j}^{\rm obs}-F_{\lambda_j}^{\vec{x}}}{\sigma}\right)^2\,.
\end{equation}
Here the sum $j$ is over the $N_{\lambda}$ wavelength points in the observed profile, with the model profile interpolated on this wavelength grid, $\sigma$ is the error in the observed relative flux (defined as inverse of the observation's S/N ratio), and $\nu=N_\lambda-5$ is the number of degrees of freedom in the fit. Thus a $\chi_\nu^2$ value is found for each combination of model parameters, $\vec{x}$, and the minimum is defined as the best-fit model. Of course, as shown below, many combinations of parameters will fit the observed line nearly as well as the best-fit one. Where possible, we consider the set of all $\{\vec{x}\}$ that satisfy $\chi_\nu^2(\vec{x})\le 2$ as being an acceptable fit and report the mean and standard deviation of $\{\vec{x}\}$ as ``the" best-fit parameters with their associated $1\sigma$ errors.

\begin{figure}
\centering
\includegraphics[width=1.0\columnwidth]{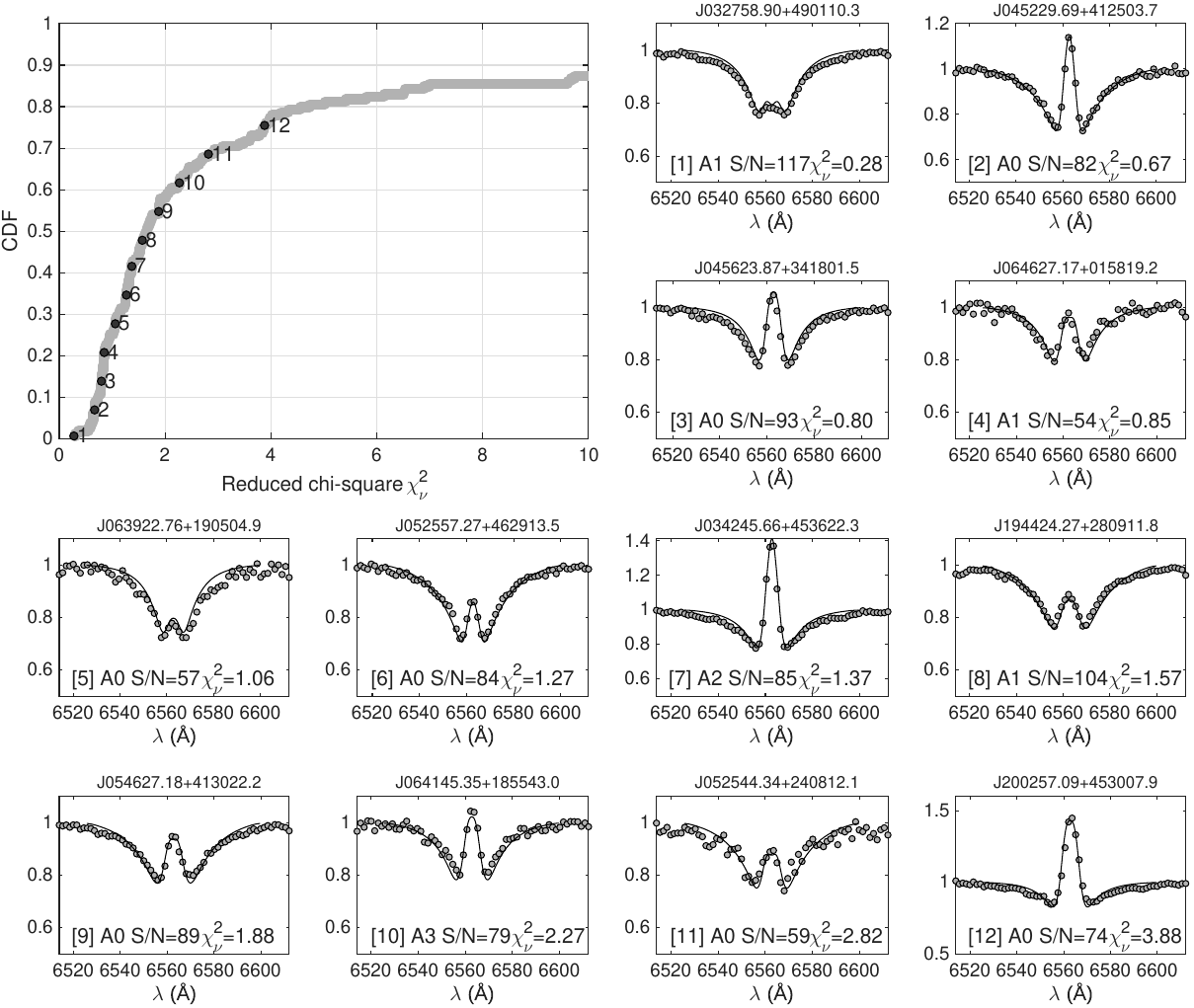}
\caption{Top-left: Cumulative distribution of the reduced-$\chi^2$ statistic for the 159 CAe sample stars. Numbered points (1-12) indicate the locations of the representative \ha\ fits shown in the surrounding panels. Panels [1]-[12]: Each compares the best-fit synthetic profile (line) with the observed spectrum (dots). The star's spectral type, the S/N ratio of the spectrum, and the $\chi_\nu^2$ of the fit are as indicated. The title of each plot is the star's LAMOST identification.}
\label{fig:Example_HalphaFits}
\end{figure}

Example fits are shown in Figure~\ref{fig:Example_HalphaFits}, which presents twelve representative profiles spanning a range of reduced chi-squared values, $\chi_\nu^2$ from 0.3 to 3.9, as well as the cumulative distribution of $\chi_\nu^2$ for the sample. The observations present a wide range of profile morphologies and emission strengths. The $\chi_\nu^2$ values of each of the twelve shown profiles are also indicated on CDF plot. For example, profile~8 (LAMOST target J194424.27+280911.8) is very close to the median reduced-$\chi^2$ of $1.6$. This indicates that the majority of CAe \ha\ profiles are well described by the axisymmetric, Keplerian disk models implemented in \texttt{Bedisk/Beray}, consistent with prior successful applications of this framework to early-type emission stars \citep{Jones2011,SigutPatel2013,Sigut2023}. The CDF’s high-\chisq\ tail, with approximately 20\% with $\chi_\nu^2> 4$, indicates that a minority of spectra have significantly poorer \ha\ fits.

\begin{figure}
\centering
\includegraphics[width=1.0\columnwidth]{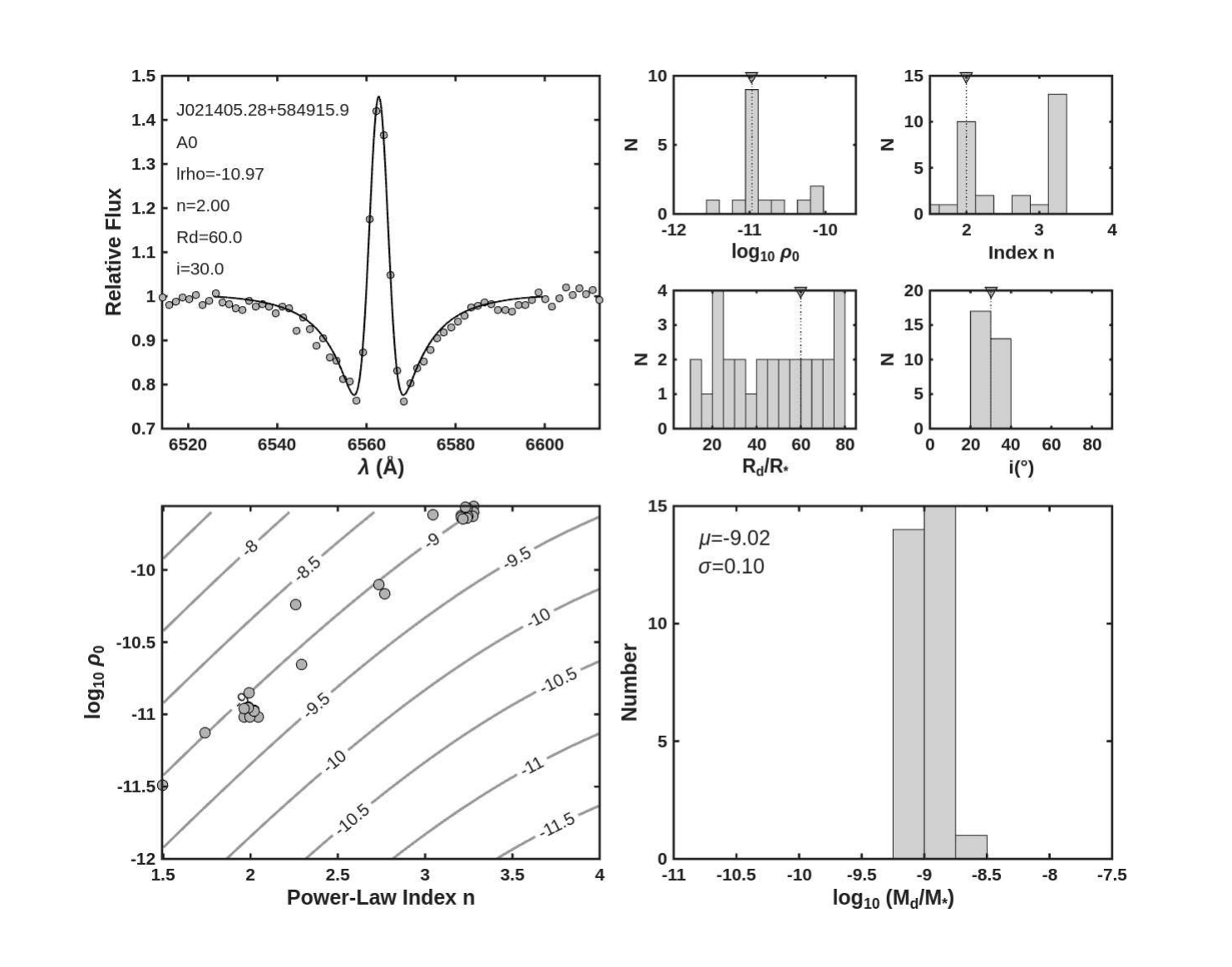}
\caption{Fit to the \ha\ line of J021405.28+584915.9. Top-left: The observed profile (circles) with the best-fit model profile (line). Top-right (4 sub-panels): Histograms of the $\rho_0$, $n$, $R_{\rm disk}$ and $i$ parameters for the 30 models that satisfy $\chi_{\nu}^2\le 2$. Parameters for the best fit profile are indicated by the arrows. Bottom-left: The $(\rho_0,n)$ parameters (circles) of the 30 models with $\chi_{\nu}^2\le 2$. Contours are lines of constant $\log(M_{\rm disk}/M_*)$ as computed by Eq.~(\ref{eq:diskmass}). Bottom-right: Histogram of the 30 $\log(M_{\rm disk}/M_*)$ estimates; the mean and standard deviation of the distribution are as indicated.}
\label{fig:HalphaPlane}
\end{figure}

Figure~\ref{fig:HalphaPlane} shows in more detail the fitting procedure for a representative object, J021405.28+584915.9, an A0 CAe star from the sample that shows a prominent, single-peak emission in the core of the wider photospheric \ha\ absorption line. This profile is best-fit by the model
$\vec{x}=(\rho_0,n,R_{\rm disk},i;\log(g))=(1.07\times10^{-11}{\rm g\,cm^{-3}},2.0,60R_*,30^\circ;4.0)$ with $\chi_\nu^2=0.82$. The panels to the right of the line profile fit show the distribution of the parameters $(\rho_0,n,R_{\rm disk},i)$ for the 30 models that fit within $\chi_\nu^2\le 2.00$. Aside from the viewing inclination that is robustly recovered as $i=30^\circ \pm 10^\circ$, the disk parameters $(n,R_{\rm disk})$ have rather broad, flat distributions leading to large formal uncertainties. For this reason, it is not meaningful to adopt the $(\rho_0,n,R_{\rm disk})$ parameters of the best-fit model as representing the disk density structure. However, it is very interesting to look at the distribution of the 30 $(\rho_0,n)$ values of the best-fitting profiles in the $\log\rho_0-n$ plane, as done in the lower-left panel of Figure~\ref{fig:HalphaPlane}. Also plotted in the $\log\rho_0-n$ plane are lines of constant disk mass (from Eq.~\ref{eq:diskmass}) assuming $R_{\rm disk}=25\,R_*$. The 30 individual pairs of $(\rho_0,n)$ that best fit the \ha\ profile cluster around the contour $\log_{10}(M_{\rm disk}/M_*)=-9.0$; the actual distribution\footnote{Using the $R_{\rm disk}$ value of the fit, not the value of $R_{\rm disk}=25\,R_*$ used for the contours in the lower-left panel of Figure~\ref{fig:HalphaPlane}.} of disk masses is shown in the lower-right panel of the figure. Thus while the individual $(\rho_0,n)$ parameters are quite uncertain, they collectively define a meaningful result for the disk mass.

\begin{figure}
\includegraphics[width=1.0\columnwidth]{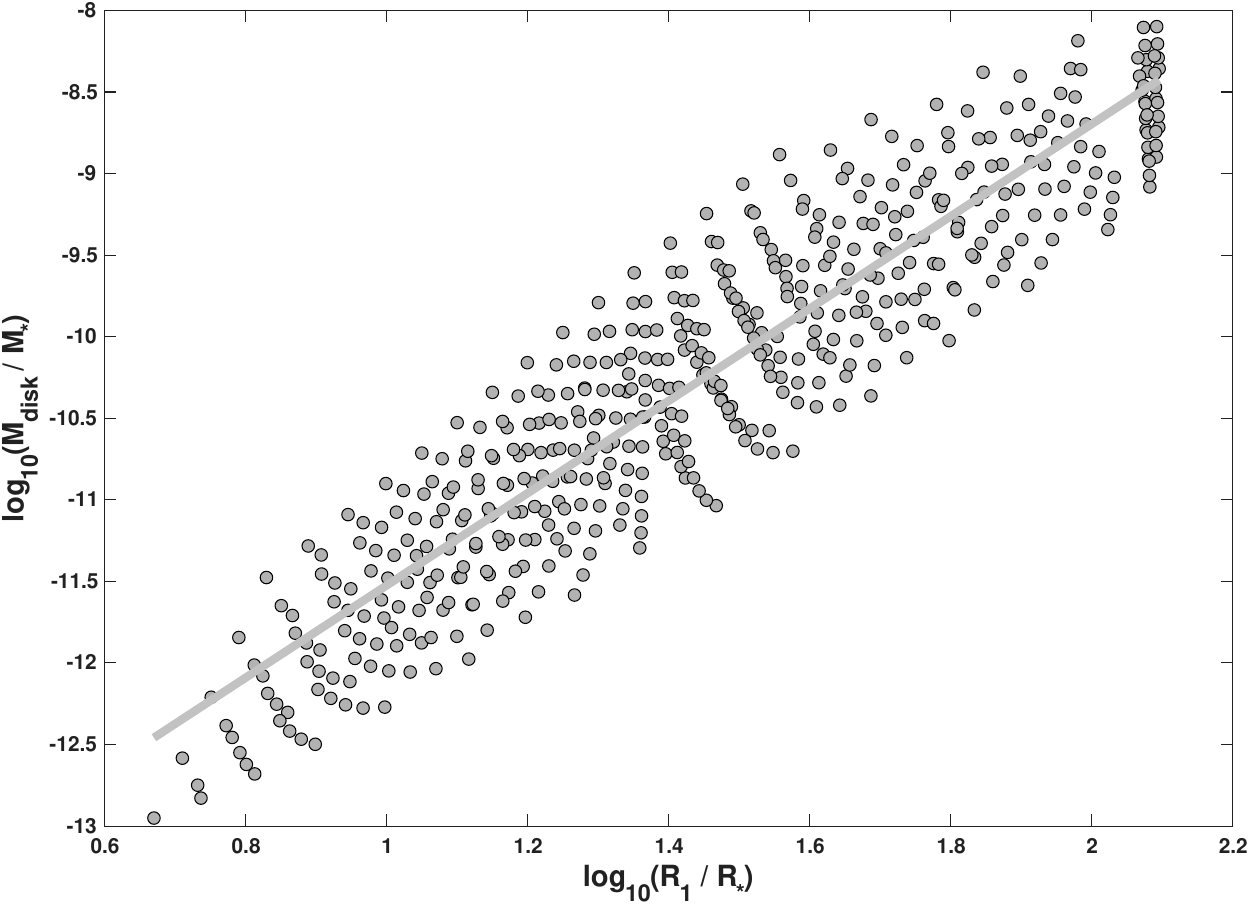}
\caption{Relation between disk mass $\log_{10}(M_{\rm disk}/M_*)$ and \ha\ radius $\log_{10}(R_1/R_*)$ defined via $\tau_0(R_1)=1$ in Eq.~(\ref{eq:hopac}). The gray line is the best-fit straight line with a slope of $2.8$.}
\label{fig:MassTau}
\end{figure}

It might seem strange that the disk mass can be constrained as optically thick lines such as \ha\ trace emitting area, not emitting volume; however, the profile fits for \ha\ constrain the disk-density parameters $(\rho_0,n,R_{\rm disk})$ that fix the disk mass according to Eq.~(\ref{eq:diskmass}). Another way to think about this is that to first order, the emission flux in \ha\ is set by the disk radius at which the line centre optical depth in \ha\ drops to one, defining the emitting surface area for \ha. The optical depth at line centre in \ha\ ($\nu=\nu_0$), vertically through the disk at distance $R$, is given by
\begin{equation}
\label{eq:hopac}
\tau_{\nu_0}(R)= \frac{h\nu_0}{4\pi}\,\int_{-\infty}^{+\infty}\,\left[N_2(R,Z) B_{23}-N_3(R,Z) B_{32}\right]\,\phi_{\nu_0}\,dZ\;.
\end{equation}
Here, only the line opacity due to \ha\ is included, utilizing hydrogen level populations $N_2(R,Z)$ and $N_3(R,Z)$ following from the \texttt{Bedisk} solution; $\phi_{\nu_0}$ is the line profile evaluated at line center, and $B_{23}$ and $B_{32}$ are the usual Einstein coefficients for the \ha\ transition. The emitting surface area of the disk for \ha\ is then from $R=R_*$ to $R=R_1$ defined by $\tau_0(R_1)=1$ in the above equation. Technically this is true only for observers at $i=0^{\circ}$ (a face-on disk), but this is sufficient for the argument we wish to make. Figure~\ref{fig:MassTau} shows the correlation between $\log_{10}(R_1)$ and $\log_{10}(M_{\rm disk}/M_*)$ for a wide sample of \texttt{Bedisk} models. Clearly these quantities are strongly correlated ($r^2=0.85$) and the trend is very well fit by a straight-line of slope $\sim 2.8$. Thus, constraining the emitting area of the disk also acts to constrain the total mass in the disk via the adopted density model.

\subsection{Distribution of Disk Masses}
\label{sec:dmass}

In this section, we look at the distribution of the disk masses found for our LAMOST 159 CAe star sample, and compare the average disk mass as a function of spectral type to those derived for a sample of late-type CBe stars derived using similar methods \citep{Arcos2017}. A complete list of disk masses for our sample stars can be found in Table~\ref{tab:appendix} in the Appendix. The total disk mass is a key parameter characterizing the circumstellar environment, directly tied to the ability of the central star to transport angular momentum and mass into a disk \citep{Granada2013a}. 

\begin{figure}
\centering
\includegraphics[width=1.0\columnwidth]{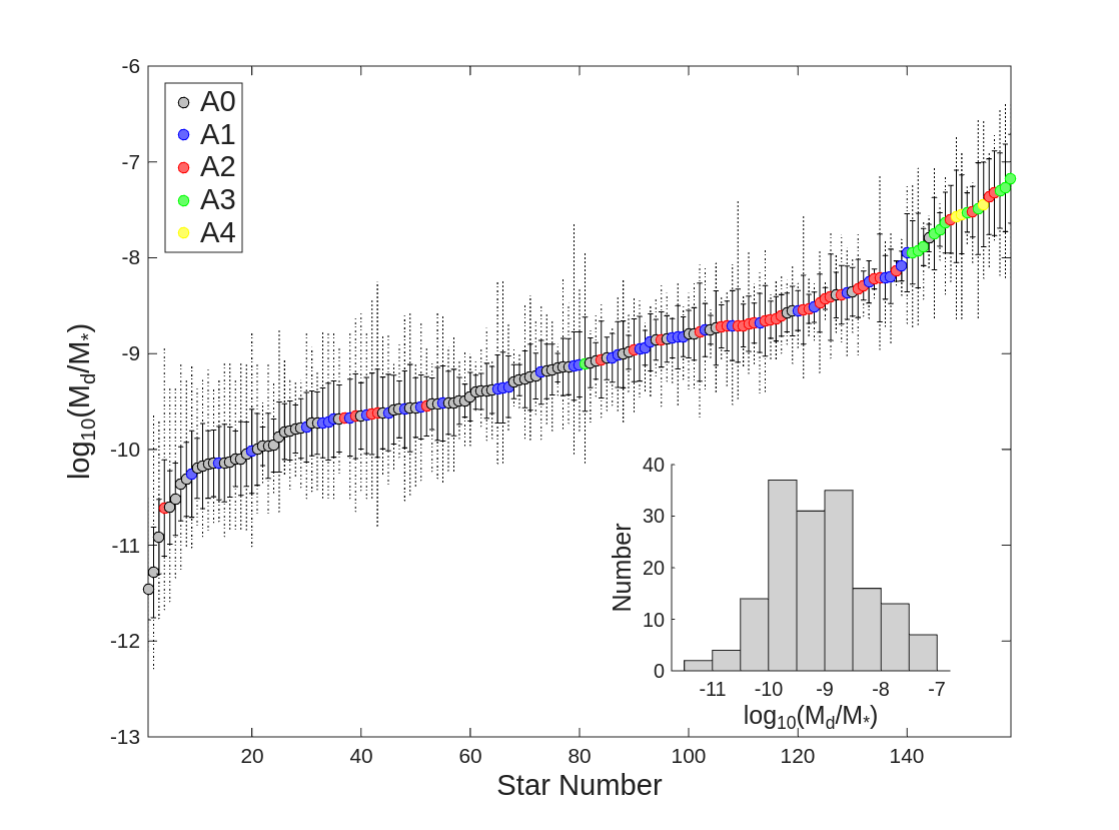}
\caption{Main: Disk masses $(\log_{10} M_{\rm disk}/M_*)$ for each sample star versus star number ordered by increasing disk mass. The coloured circles are the mean of the disk masses obtained from the closest-fitting profiles and the error bars are $1\sigma$. The dotted lines connect minimum and maximum disk mass. The symbol colours identify the star's spectral type. Insert: histogram of $\log_{10} (M_{\rm disk}/M_*)$ for the sample. The distribution is consistent with a Gaussian of mean $\mu=-9.07$ and standard deviation of $\sigma=0.85$. }
\label{fig:Dmass_Curve}
\end{figure}

Figure~\ref{fig:Dmass_Curve} shows the disk mass derived from the \ha\ profile fits for all 159 stars in the LAMOST sample, ordered by increasing disk mass. The error bars for each star are $1\sigma$ as discussed in the previous section (see also Figure~\ref{fig:HalphaPlane}). The figure insert shows a histogram of $\log_{10}({M_{\rm disk}/M_*})$, and a KS test to the distribution indicates that it is normally-distributed with a mean (over all spectral types, A0 through A4) of $\mu=-9.1$ and a standard deviation of $\sigma=0.85$. Also clear from the main figure is that the latest spectral types (A3 and A4) are preferentially found at large disk masses with $\log({M_{\rm disk}/M_*})\ge -8$, a point we return to below.

\begin{figure}
\centering
\includegraphics[width=1.0\columnwidth]{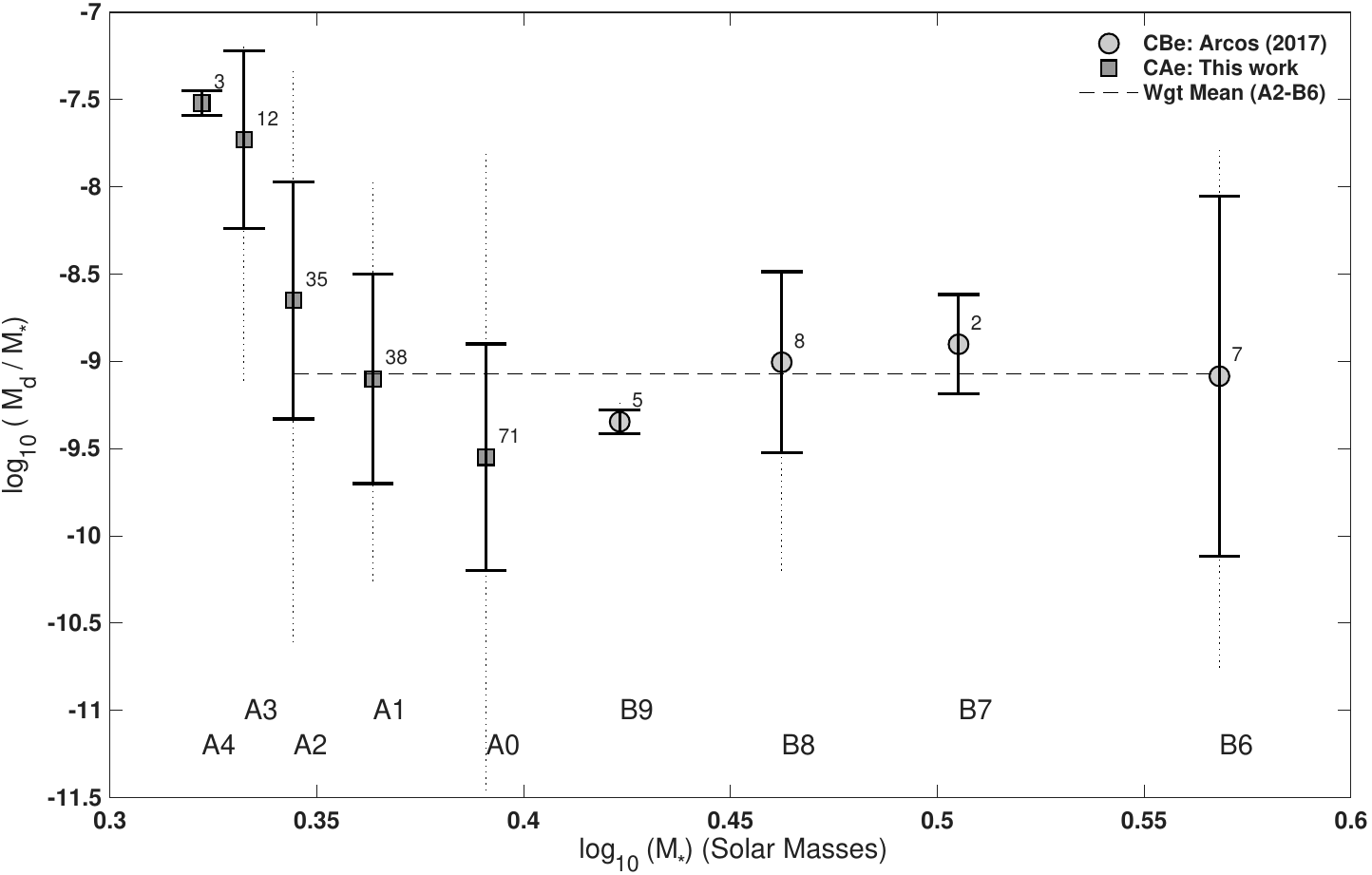}
\label{fig:disktrend}
\caption{The average $\log_{10}(M_{\rm disk}/M_*)$ as a function of stellar mass $\log_{10}({M_*/M_\odot})$. The solid error bars are $1\sigma$ and the dotted lines connect the smallest and largest disk mass at that spectral type. The squares are CAe disk masses from the current work and the circles are disk masses for late-type CBe stars from \citet{Arcos2017}. The number of stars in each spectral type sample is given next to the symbol. The dashed line at $\log_{10}(M_{\rm disk}/M_*)=-9.1$ is the average disk mass in the spectral type range A2 to B6 (see text).}
\label{fig:Dmass_SpecType}
\end{figure}

Figure~\ref{fig:Dmass_SpecType} plots the average disk masses as a function of spectral type, along with the standard deviation and minimum and maximum masses. For a broader context, we have also included the disk masses estimated (using similar methods based on \texttt{Bedisk/Beray}) for late-type CBe stars from \citet{Arcos2017}\footnote{The sample sizes for the CBe stars are smaller than in the present work; however, the underlying \ha\ spectra, taken from the BeSOS survey (besos.ifa.uv.cl), are of much higher resolution.}. Treating spectral types A3 and A4 as outliers to be omitted (for reasons discussed below), we perform a Monte Carlo simulation in which the average disk masses at each spectral type are randomly realized within their errors (assuming a Gaussian distribution) and then a straight line is fit to the $\log_{10}M_*$ versus $\log_{10}({M_{\rm disk}/M_*})$ relation. Five hundred repetitions show that the slope of this line is $-0.03 \pm 4.2$, consistent with a constant value over this spectral type range. A weighted-average $(w=1/\sigma)$ gives $<\!\log({M_{\rm disk}/M_*})\!>=-9.1$. This figure indicates that there is a continuity in disk mass from the late-type CBe stars into the CAe star range.  

\begin{figure}
\centering
\includegraphics[width=1.0\columnwidth]{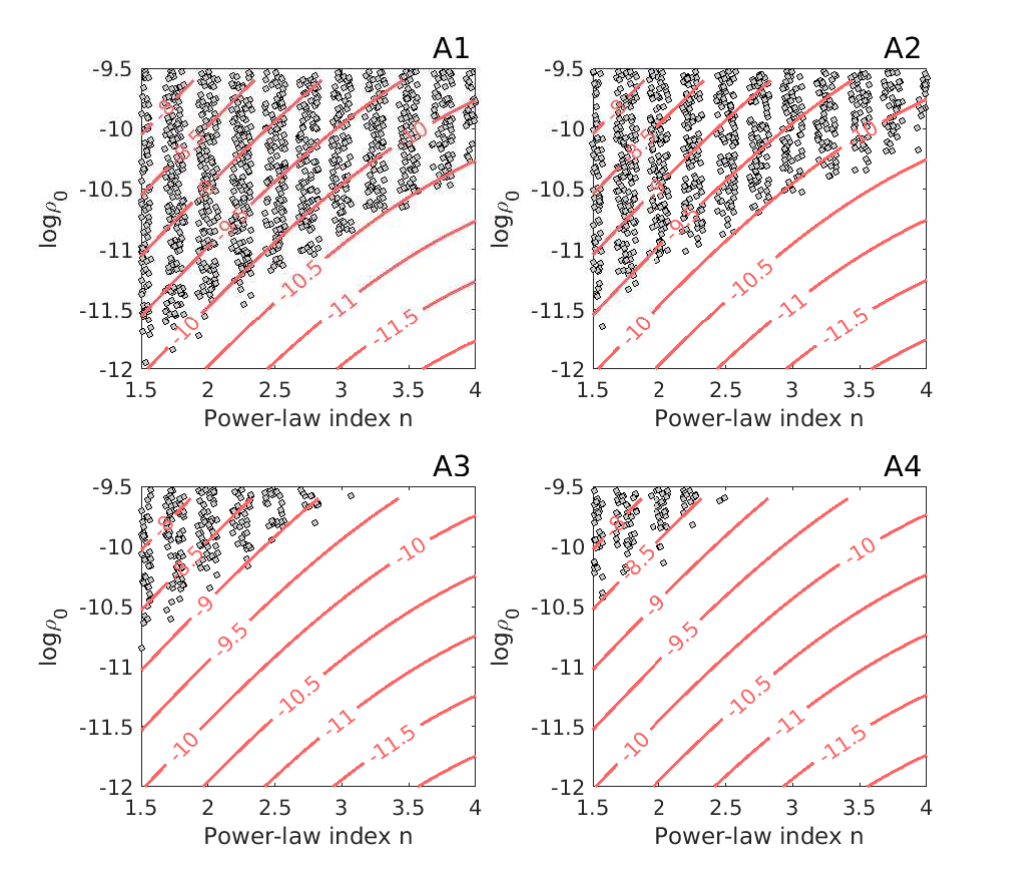}
\caption{Each panel shows for the indicated spectral type, the combinations of $\log(\rho_0),n$ that result in detectable \ha\ emission profiles with $i\ge 25^\circ$. Detectability is based on \ha\ profile morphology as discussed in the text. The contours in each plot are curves of constant $\log_{10}(M_{\rm disk}/M_*)$. The spectral type A0 is not shown as its distribution is very similar to the spectral type A1.}
\label{fig:Emission_Plane}
\end{figure}

As noted above, we have omitted spectral types A3 and A4 as outliers in Figure~\ref{fig:Dmass_SpecType}. There is good evidence that the rise in disk mass for these late spectral types is systematic shift in the spectroscopic detectability of a disks using \ha\ for the coolest spectral subtypes. Figure~\ref{fig:Emission_Plane} illustrates the detectability of circumstellar disks using \ha\ as a function of the base disk density $\rho_0$ and the density exponent $n$. The four panels show spectral subtypes A1 through A4 separately (A0 stars are not shown as their behaviour is nearly identical to A1). Over-plotted are contours of constant disk mass, $\log_{10}(M_{\rm disk}/M_*)$, ranging from $-12$ to $-8$. Each symbol indicates a model that produces detectable \ha\ emission, with slight jitter added for clarity. To define a detectable disk, we have classified all computed \ha\ lines by morphological type: single-peaked emission, double-peaked emission, or pure absorption (i.e.\ a photospheric profile), and thus were able to eliminate $(\rho_0,n)$  combinations that produced only photospheric profiles. As noise was not considered, this is a very conservative definition of detectability, as even weak emission is counted. The full range of model parameters $(\rho_0,n,R_{disk},i)$ were considered at each spectral type except the viewing inclination, which was constrained to be $i \geq 25^{\circ}$. The inclination was constrained in this way to reflect the low probability of such viewing inclinations ($\le 10$\%) occurring in small samples sizes, such as the case in our A3 and A4 samples.\footnote{Small viewing inclinations often place a small emission peak near line centre in \ha, resulting in the model flagged as detectable. In practice, low inclinations are improbable due to the assumed $\sin i$ distribution of random viewing inclinations. In addition, the inclusion of a finite (S/N) would also render weak emission undetectable.} 

A systematic progression with spectral type is evident. For spectral type A1 and A2 stars (and A0 stars, not shown) detectable disks occupy a broad range of density structures, corresponding to relatively low detectable disk masses in the range $\log_{10}(M_{\rm disk}/M_*) \lesssim -10.5\;({\rm A1})$ or $-10.0\;({\rm A2})$. In sharp contrast, the A3 and A4 stars exhibit minimum detectable disk masses of $\log_{10}(M_{\rm disk}/M_*) \ge -9\;({\rm A3})$ or $-8.5\;({\rm A4})$. Hence the upward trend in disk mass for types A3 and A4 seen in Figures~\ref{fig:Dmass_Curve} and \ref{fig:Dmass_SpecType} almost certainly reflects this systematic trend that only the densest and most massive disks would be detectable in \ha.

In conclusion, we find direct evidence that the CAe stars have disk masses very similar to the disk masses of late-type (B6 and later) CBe stars. The CAe and CBe stars in the spectral type range (A2 through B6) are all characterized by an average \ha\ disk mass of $\log_{10}(M_{\rm disk}/M_*) = -9.1$. This trend may continue to even later-type CAe stars, but this characteristic disk mass becomes undetectable in \ha\ for spectral types A3 and later. 

\subsection{Test for Random Inclinations}
\label{sec:inclination}

\begin{figure}
\centering
\includegraphics[width=1.0\columnwidth]{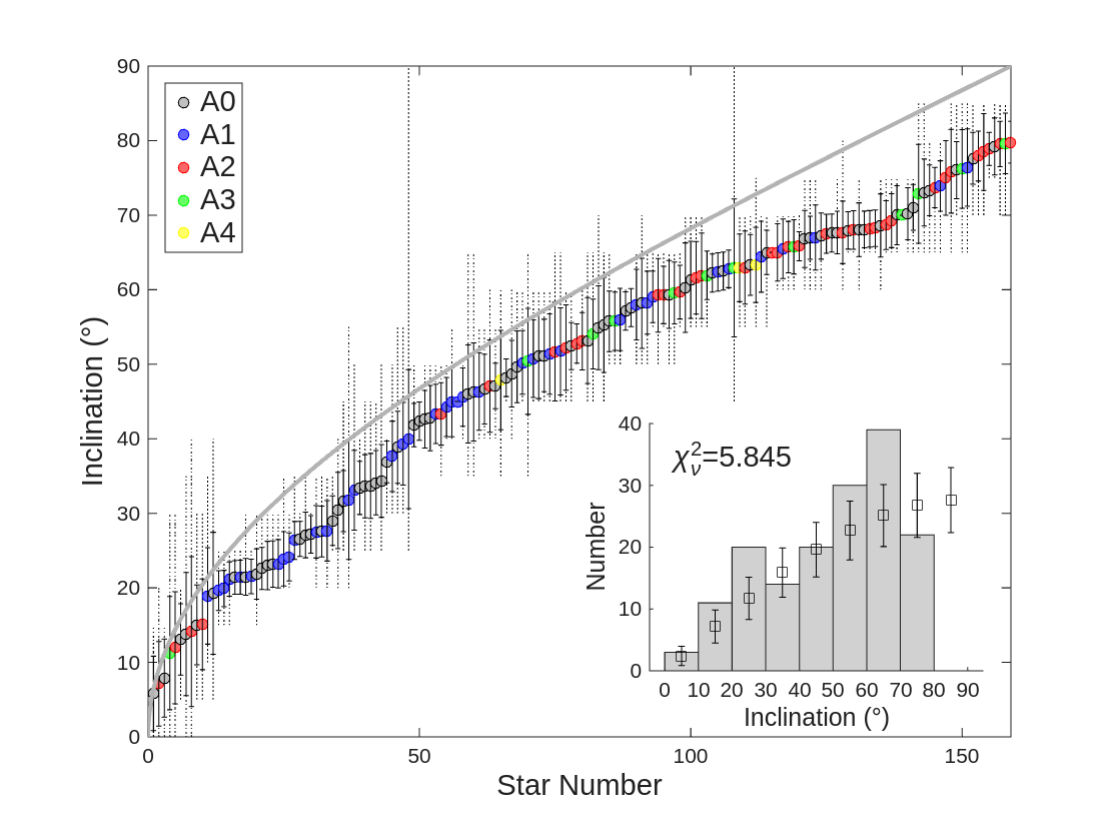}
\caption{Main: Inclination angle $i$ versus star number for the CAe sample sorted by inclination. Each star's inclination (circle) is the mean of the inclinations of the closest-fit profiles and the errors bars are $1\sigma$. The dotted line connects the minimum and maximum inclinations. The solid curve is the expected relation based on the $\sin i$ distribution. The symbol colours identify each star's spectral type. Insert: A histogram of the inclinations compared to the number expected based on the $\sin i$ distribution (squares). The reduced chi-squared for the comparison is $\chi_{\nu}^2=5.845$.}
\label{fig:Inc_Curve}
\end{figure}

As the CAe stars in our LAMOST sample are unassociated, one would expect the distribution of system inclination angles (the angle between the star's rotation axis and the line of sight) to follow the $p(i)di=\sin i\, di$ distribution expected for randomly-oriented stellar rotation axes \cite{Gray2022}. As the system inclination is one of the parameters determined by the fitting procedure for \ha, this expectation can be directly tested. Figure~\ref{fig:Inc_Curve} shows the derived stellar inclination angles with errors for all 159 sample stars ordered by increasing inclination; the figure insert shows a histogram of the sample inclinations, as well as the expected number from the $\sin i$ distribution. A complete list of inclination angles for our sample stars can be found in Table~\ref{tab:appendix} in the Appendix.

In Figure~\ref{fig:Inc_Curve}, the solid curve gives the number of stars in our sample expected below a given inclination based on the $\sin i$ distribution, $N(\le i) = 159\,(1-\cos i)$. This figure also illustrates the errors in the inclination estimates, taken to be the standard deviation of the inclinations corresponding to the best-fit models (see Figure~\ref{fig:HalphaPlane}), and the minimum and maximum inclination for each star. Errors are typically large for $i\le 20^\circ$ and in some individual cases depending on the quality of the profile fits.  It might seem surprising that inclinations angles (the ``$i$" in $v\sin i$) can be determined from low-resolution spectra at all. At ${\cal R}=1800$, the velocity resolution is $167\,\rm km\,s^{-1}$. For a star rotating at an equatorial velocity of $280\,\rm km\,s^{-1}$ (see Table~\ref{tab:Astars}), it's $v\sin i$ value falls below this nominal resolution for $i\le 37^{\circ}$. If the derived inclination was dependent solely on photospheric line widths, inclination angles below $37^{\circ}$ could not be distinguished. However, in the current case, the inclination is derived mainly from morphology of the \ha\ emission line profile \citep{Porter2003}. As illustrated by Figure~1 of \citet{Lailey2024}, the \ha\ morphology changes from singly-peaked emission for low $i$, to doubly-peaked emission for moderate $i$, and finally to strong shell absorption for high $i\ge 80^{\circ}$. Thus the \ha\ fitting method of \cite{Sigut2020} and \cite{Sigut2023} can detect low inclinations, much lower than that expected based on photospheric line widths alone.

One notable feature of Figure~\ref{fig:Inc_Curve} is that the observed number of stars slightly exceeds expectations based on $\sin i$ for $i\le 30^\circ$, and there is a complete lack of higher inclinations for $i\ge 80^\circ$. This is made more clear in the figure insert which shows a histogram of sample inclinations along with the number predicted by the random $\sin i$ distribution. The deficit of high inclination objects is now noticeable for $i\ge 70^\circ$ and becomes strongly evident for $i\ge 80^\circ$, where no observed sample stars are seen. A chi-squared test gives $\chi_{\nu}^2=5.85$, and the null hypothesis that the observed sample is consistent with random inclinations is rejected.

Our lack of high-inclination CAe stars is very reminiscent of the situation for the CBe stars, where this issue was first noted by \citep{Rivinius2006a} and is reflected in samples of inclinations derived from gravitational darkening of photospheric line profiles \citep{Zorec2016} and \ha\ line profile morphology \citep{Sigut2023}. The key insight is that for high inclination, the disk is seen nearly edge on and most rays through the disk trace back to the stellar photosphere. In this case, in addition to emission, deep shell absorption appears when the line centre flux in \ha\ often drops below the underlying photospheric profile. Spectroscopic searches for CBe stars usually focus on emission signatures in \ha\ and can be biased against identifying high inclination objects, as directly demonstrated by \citet{Lailey2026}. The situation for the CAe stars may be even more extreme compared to the CBe stars as CAe \ha\ emission is weaker; shell absorption may be accompanied only by very weak emission outside core which is easily missed by automated detection algorithms. In conclusion, The lack of high-inclination CAe stars likely reflects observational biases rather than an intrinsic property of the population.

\section{\texorpdfstring{Ca\,\textsc{ii} IR Triplet Emission}{Ca II IR Triplet Emission}}
\label{sec:ca-ii}

The \caII\ infrared triplet (IRT; $\rm 4p\,^3P^o\rightarrow 3d\,^3D$; $\lambda\lambda\,8498,8542,8662\;$\AA) is commonly observed in emission in CBe stars and is associated with dense circumstellar gas in the inner regions of decretion disks \citep{Polidan1976, Andrillat1990}. Observational studies have shown that the \caII\ IRT strength correlates with \ha\ emission and exhibits similar variability patterns, suggesting that both diagnostics originate from dynamically evolving gaseous disks \citep{Koubsky2012, Banerjee2021}. Further investigations indicate that \caII\ emission forms in relatively cool and dense inner CBe disk layers, making it particularly sensitive to the local density structure and ionization conditions of the circumstellar environment \citep{Apparao1988, Shokry2018}.

The presence of \caII\ IRT emission in CAe stars supports the idea that CAe and CBe stars form a continuous sequence of rapidly rotating emission-line stars with structurally similar decretion disks. Despite its diagnostic potential, systematic modelling of \caII\ IRT emission in CAe stars has remained largely unexplored. In this section, we use the \texttt{Bedisk/Beray} codes to computed theoretical line profiles for the \caII\ IRT using the same disk density models as for \ha\ in the previous sections. We note that the Ca\,{\sc ii} IRT triplet has been modelled for Herbig Be (HBe) stars \citep[see][]{Patel2016,Patel2017}.

\begin{figure}
\centering
\includegraphics[width=0.99\columnwidth]{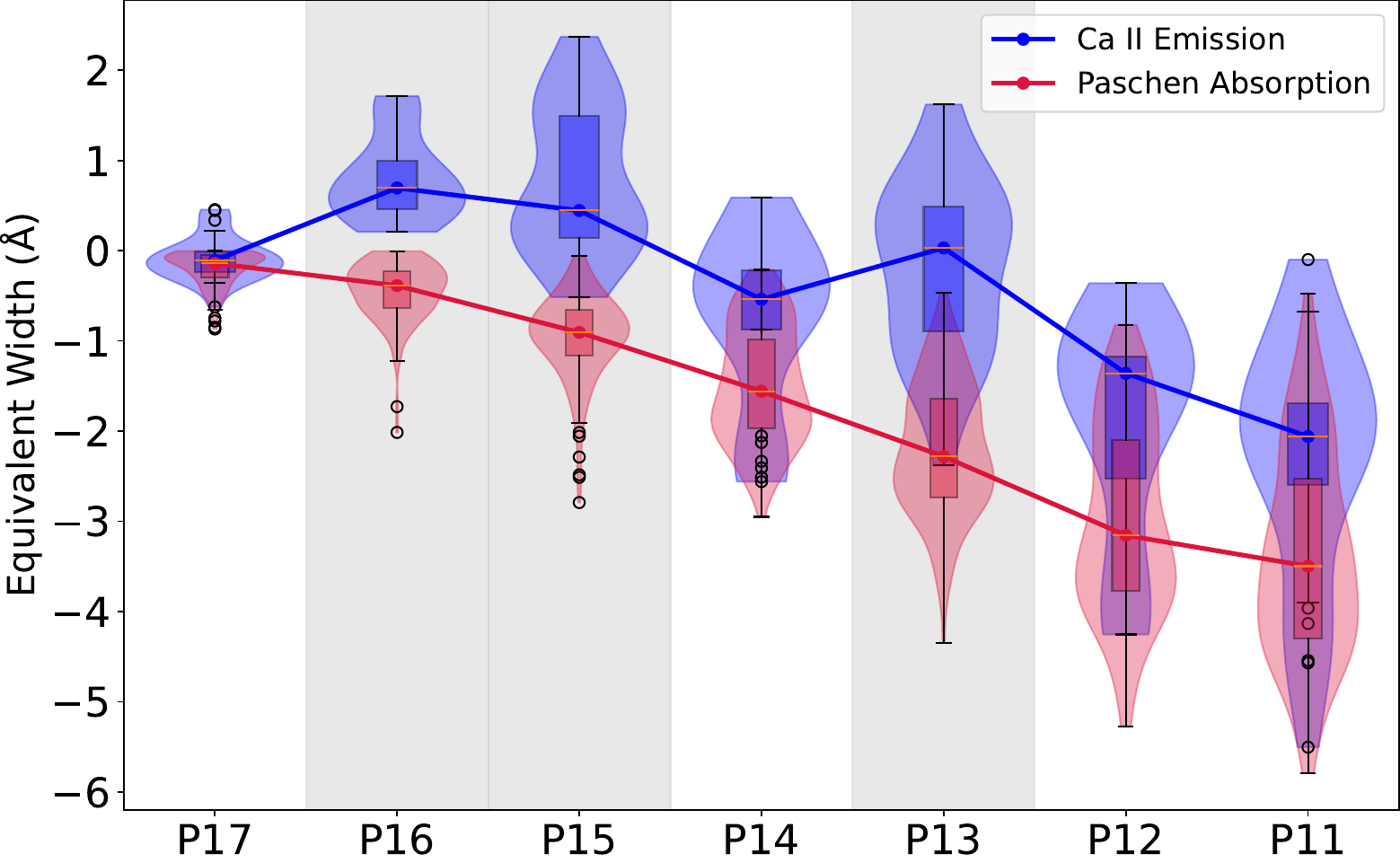}
\caption{Equivalent width (EW) distributions for Paschen lines P11 through P17 in spectra visually classified as \caII\ IRT emitters (blue) and non-emitters (crimson). 
Violin profiles show the kernel-density distribution of EW measurements, while boxplots represent the interquartile range (IQR: 25th-75th percentile); whiskers denote the remaining data extent and circles indicate outliers beyond 1.5~$\times$~IQR. The lines are smooth fits to the median values. Positive EW values correspond to net emission and negative values to net absorption. 
}
\label{fig:CaIIvsPasObs}
\end{figure}

In our LAMOST sample of 159 CAe stars, there are 25 ($16\,\%$) with detectable \caII\ emission, spanning the spectral range A0 through A3. The distribution is strongly dominated by A0 stars; of the 25, only two are A1, five are A2, and one A3; no A4 stars with \caII\ emission were identified. A complexity when dealing with the \caII\ IRT in CBe/CAe systems is that all three lines are close blends with the high-n Paschen series of hydrogen: $\lambda\,8498$ with P16 ($\lambda\,8502$),  $\lambda\,8542$ with P15 ($\lambda\,8545$), and $\lambda\,8662$ with P13 ($\lambda\,8665$). Thus, \caII\ emission is identified as anomalous emission in the Paschen series as illustrated in Figure~\ref{fig:CaIIvsPasObs}. Shown are the measured equivalent widths of the Paschen series, P11 through P17 for our CAe sample stars. The 25 stars with identified \caII\ IRT emission are clearly reflected as enhanced emission at P13, P15 and P16 compared to the other members of the sample. This figure also makes it clear that it is difficult to identify weak \caII\ emission because of the Paschen blending, particularly at low spectral resolution.   

\begin{figure}
    \centering
    \includegraphics[width=1.0\linewidth]{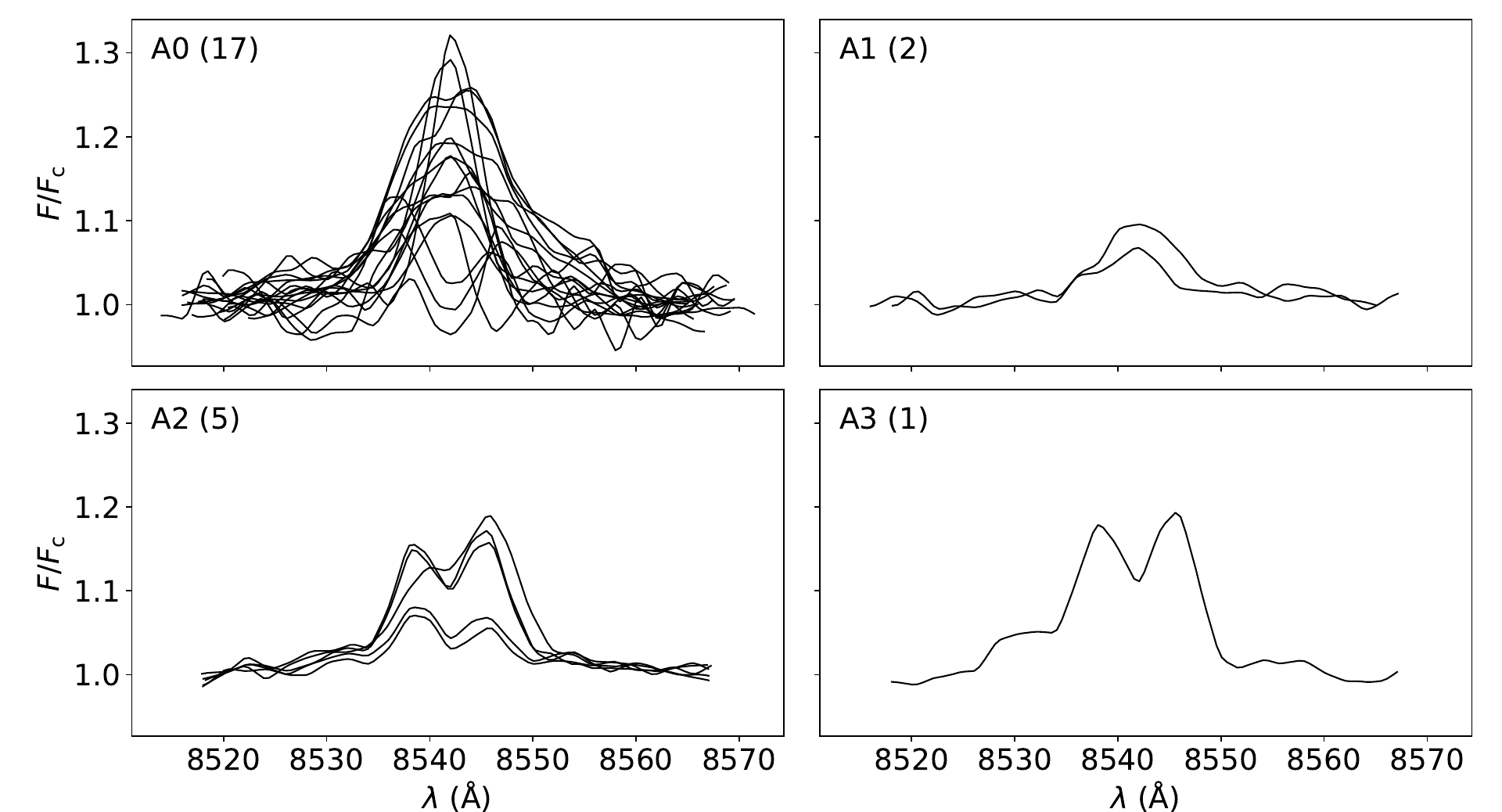}
    \caption{Observed Ca\,{\sc ii} $\lambda\,8542$ profiles for the 25 sample CAe stars with detectable emission, corrected for P15 emission, and organized by spectral type. The number of stars at each spectral type is given in brackets.}
    \label{fig:CaIIObs}
\end{figure}

Among the three IRT multiplet components, $\lambda\,8542$, $j=3/2\rightarrow 5/2$, was selected for quantitative modelling as it is generally the strongest member. Figure~\ref{fig:CaIIObs} shows $\lambda\,8542$ for all 25 sample stars with \caII\ emission, organized by spectral type. Among these objects, 13 show single-peaked emission and 12 exhibit double-peaked emission. Because of the low resolution of our LAMOST spectra, the observed profiles of the Ca\,{\sc ii} IRT lines mainly reflect the instrument profile, and the main observables that can be extracted are the line equivalent width ($EW$) and the peak flux ($F_p$) compared to the continuum. We made a simple correction for the hydrogen P15 blend by subtracting (if detected) the unblended hydrogen P14 profile lying between $\lambda\,8542$, and $\lambda\,8662$ from the composite $\lambda\,8542$-P15 profile, a procedure similar to \citet{Patel2016,Patel2017}. We then measured the peak flux and equivalent width from the subtracted profile. In almost all cases, these corrections were not large, particularly for the peak flux. Errors in these quantities were estimated from the spectrum (S/N) and the uncertainty in the continuum placement, but do not reflect uncertainties associated with the subtraction process.

\begin{figure}
\centering
\includegraphics[width=1.0\columnwidth]{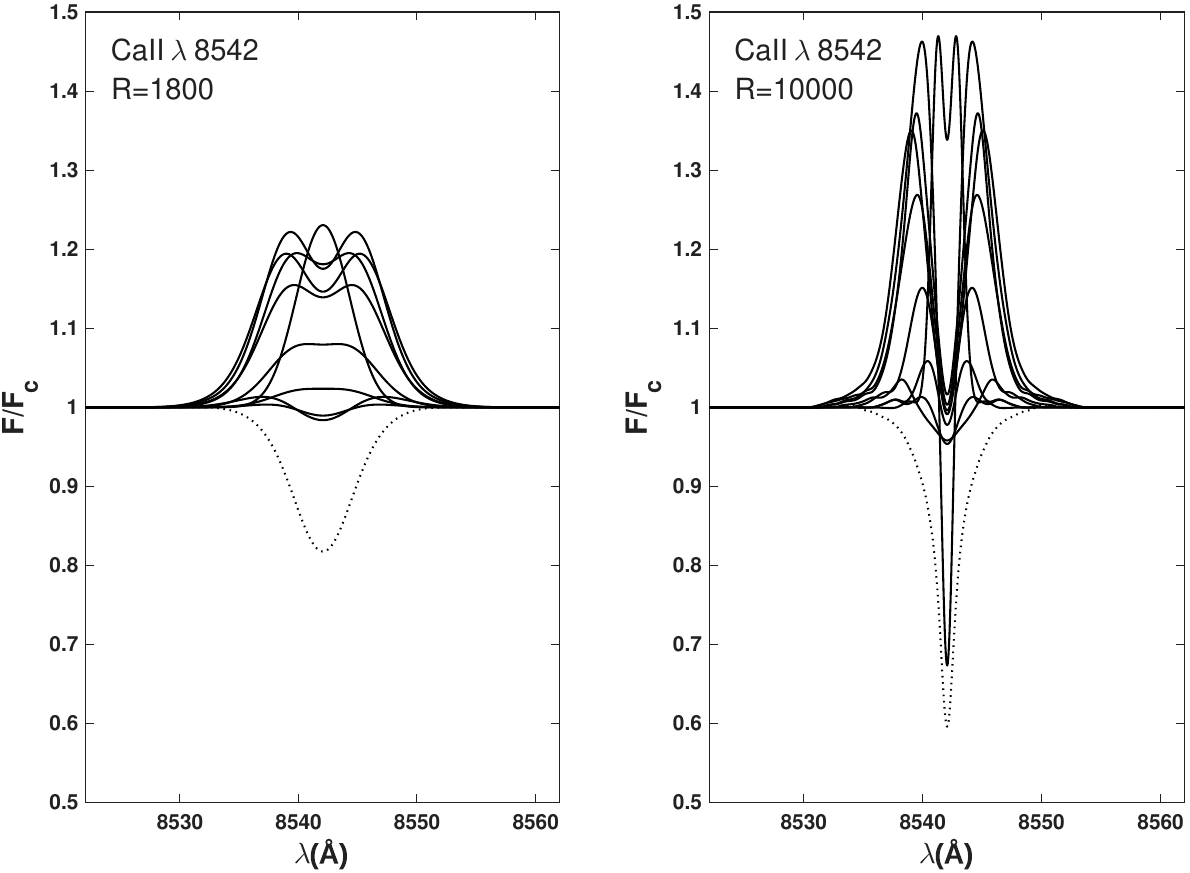}
\caption{Ten example computed profiles for Ca\,{\sc ii} $\lambda\,8542$, seen at spectral resolution ${\cal R}=1800$ (left) and ${\cal R}=10,000$ (right). The dotted profile in both panels is the photospheric \caII\ absorption of the central star.}
\label{fig:CaIIExamples}
\end{figure}

We compare the $(EW,F_p)$ distribution of Ca\,{\sc ii} $\lambda\,8542$ with \texttt{Bedisk/Beray} predictions over the same range of parameters considered previously for \ha. \texttt{Bedisk} provides the $\rm 4p\,^3P^o_{3/2}$ (upper) and $\rm 3d\,^3D_{5/2}$ level populations from the thermal solution, and \texttt{Beray} performs the formal solution using an $A_{ji}$ value\footnote{Taken from the NIST spectral lines database at physics.nist.gov/PhysRefData/ASD$\;$.} of $9.9\times 10^6\;\rm s^{-1}$ and an intrinsic Doppler line profile. Example computed line profiles are shown in Figure~\ref{fig:CaIIExamples} that lie in the same $EW$ range as the observations, for both ${\cal R}=10000$ and ${\cal R}=1800$. This figure clearly illustrates how the line profiles are changed by spectral resolution. At ${\cal R}=10000$, most of the profiles exhibit double-peaked emission; however, at ${\cal R}=1800$, double-peaked emission is much less prominent. In addition, while the EW is unaffected by resolution, the peak flux is significantly reduced at lower resolution.

\begin{figure}
\centering
\includegraphics[width=1.0\columnwidth]{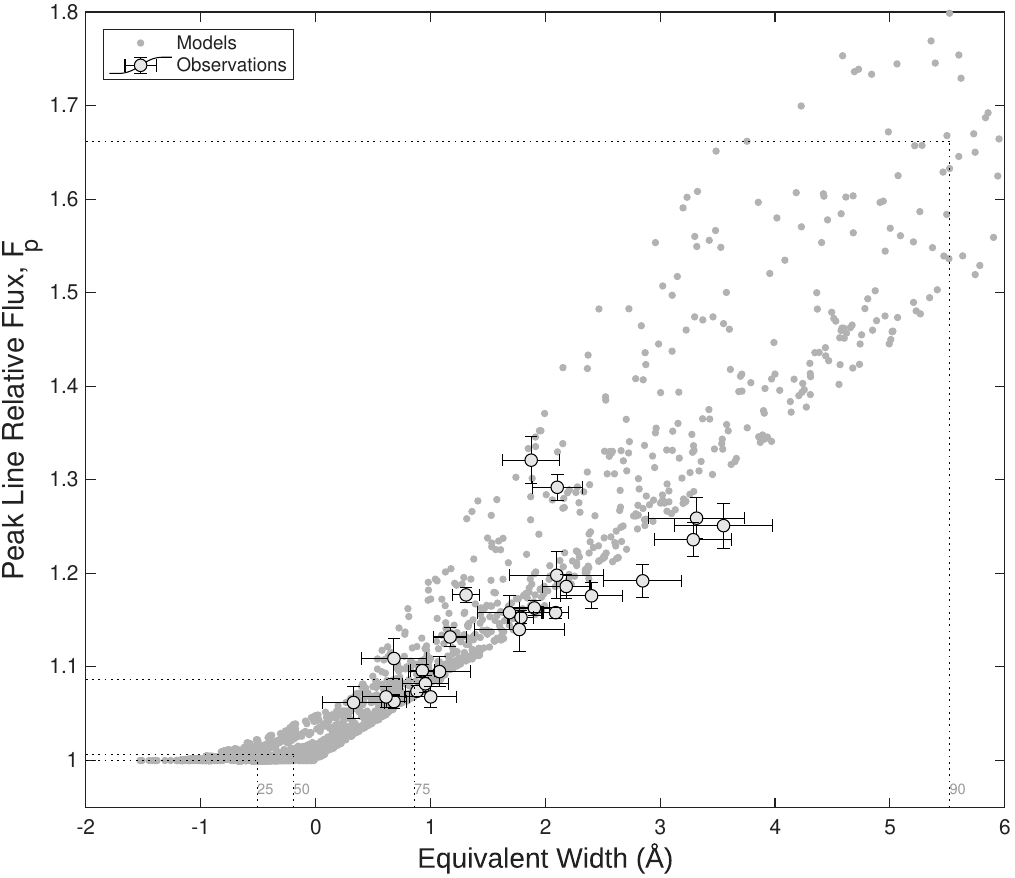}
\caption{Comparison of Ca\,{\sc ii} $\lambda\,8542$ emission between synthetic profiles (light gray points) and observed CAe stars (open circles with error bars) in the $(EW,F_p)$ plane. Regions containing 25, 50, 75, and 90\% of the models are indicated by the dotted lines.}
\label{fig:CaIIpeakEW}
\end{figure}

Figure~\ref{fig:CaIIpeakEW} compares the distribution for the observed (P15 corrected) $\lambda\,8542$ line with the model predictions for ${\cal R}=1800$ in the $(EW,F_p)$ plane. The observations are shown with estimated errors in both $EW$ and $F_p$. Also shown are the predictions of 2,000 model calculations picked randomly from the line profile library but with the inclinations chosen from the $\sin i$ distribution. Also shown in the figure are regions in the $(EW,F_p)$ plane that enclose 25, 50, 75, and 90\% of all of the models. The overlap between the observations and models is generally good, although the EWs for the observations seem to trend slightly larger than the predictions, this effect is not pronounced. It is difficult to accurately measure EW for wide lines due to continuum placement uncertainties and the EW is most affected by the details of the P15 subtraction procedure. Most of the observations lie in the region stronger than 75\% of the models but significantly less than 90\%. As noted previously, only 16\% of the observed sample shows Ca\,{\sc ii} emission. Because of the blending with the hydrogen Paschen series, weak Ca\,{\sc ii} emission would be difficult to detect because of blending; hence, the strongest Ca\,{\sc ii} emission is the most readily detected.      

\begin{figure}
\centering
\includegraphics[width=1.0\columnwidth]{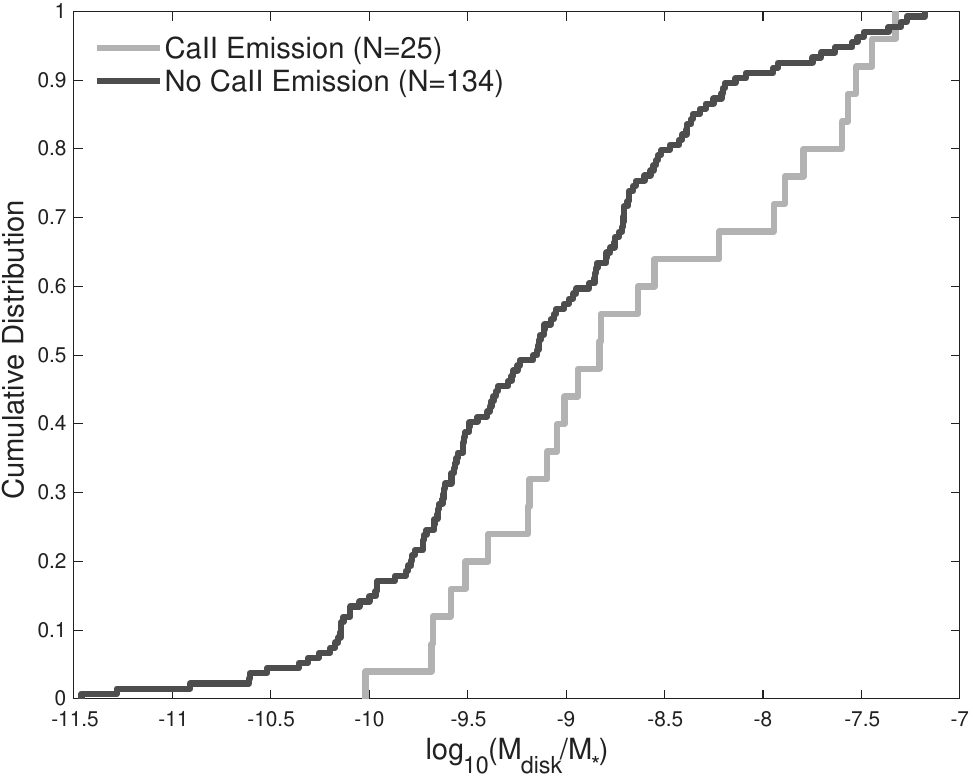}
\caption{Cumulative distribution of disk masses for sample stars with detected \caII\ IRT emission (gray line) and without detected \caII\ IRT emission (black line).}
\label{fig:CaIIDiskMass}
\end{figure}

Finally, as we have \ha\ disk masses for all sample stars, we can examine if \caII\ emission is associated with disk mass. Figure~\ref{fig:CaIIDiskMass} shows the cumulative distributions of $\log_{10}(M_{\rm disk}/M_*)$ for both the 25 objects with detected \caII\ emission and the 134 objects without detected \caII\ emission. The median disk mass $\log_{10}(M_{\rm disk}/M_*)$ is $-8.83$ for the \caII\ emitters and $-9.16$ for the non-emitters. A Wilcoxon ranked-sum test \citep{Wall2003} rejects the null hypothesis ($p=0.0086$) that the two samples come from continuous distributions with equal means, confirming the visual appearance of Figure~\ref{fig:CaIIDiskMass} that \caII\ emitters are systematically shifted to larger disk masses. However the statistical independence of the two samples, and as noted above, is a not entirely clear. Because of the blending of the \caII\ IRT and the hydrogen Paschen series, the detectability of \caII\ emission may be to some extend controlled by the hydrogen emission (and hence the \ha\ disk mass). A proper assessment of this effect would require detailed spectral synthesis for this wavelength region to allow weaker \caII\ emission to be unambiguously disentangled from the hydrogen Paschen emission.

\section{Conclusions}
\label{sec:concl}

Understanding the physical and geometric properties of circumstellar disks around CAe stars is essential for constraining their formation mechanism(s) and for situating them within the broader class of disk-bearing main-sequence stars. The CAe stars form a transitional regime between non-emission A stars and the more luminous CBe stars. In this study, we have analyzed a large sample of 159 CAe stars using synthetic \ha\ and \caII\ line profiles computed with the \texttt{Bedisk/Beray} radiative transfer models to infer disk masses, system inclinations, and disk density structure.

The inferred disk-to-stellar mass ratios, $\log_{10} (M_{\rm disk}/M_*)$, span from $-11.5$ to $-7.2$, with a sample mean of $\mu=-9.1$ and a dispersion of $\sigma=0.84$ dex. The distribution is consistent with a log-normal form, as confirmed by a Kolmogorov–Smirnov test. Comparing average CAe disk-to-star mass ratios for spectral types A0, A1, and A2 with average values found for late-type CBe stars from \citet{Arcos2017} for spectral types B6 through B9, we find that all are consistent with the single value $\log_{10} (M_{\rm disk}/M_*)=-9.1$, independent of spectral type. In this comparison, our average disk masses for spectral types A3 and A4 (with very few sample representatives) were excluded because the detectability for these latter spectral types is strongly skewed toward more massive disks. The continuity of disk mass across the CAe-CBe $T_{\rm eff}$ boundary is another strong piece of evidence that the CAe stars are the continuation of the Be phenomena into the A spectral types.

We also checked our sample for randomly-oriented rotation axes by looking for the expected $\sin i$ distribution in the recovered inclinations. While we found reasonable agreement with the $\sin i$ distribution for $i\le 70^\circ$, there was a complete absence of high-inclination systems with $i\ge 80^\circ$ where the largest fraction of systems should lie. This situation is reminiscent of the CBe stars, where a lack of high-inclination systems is well-known \citep{Rivinius2003,Zorec2016,Sigut2023} and likely due to selection effects in CBe star candidate selections \citep{Lailey2026}.

Finally, we found that the strong \caII\ IRT emission found in approximately 16\% of the observed sample is well within the predicted line strengths using disk density models that match the observed \ha\ line strengths, and that stars with detectable \caII\ emission had generally more massive \ha\ circumstellar disks.

A limitation of the low resolution spectral used here (${\cal R}=1800$) is that the disk density parameters $(\rho_0,n,R_d$) are only weakly constrained by the profile-fitting method. Profile fitting of higher-resolution spectra, including metallic lines such as \caII, may rectify this situation.

Finally, the missing high-inclination systems ($i\ge 80^\circ$) likely manifest as A-type shell stars, and systematic identification of these objects in large surveys such LAMOST may help to fill in an important, but missing portion of the CAe population.

\begin{acknowledgments}
The authors wish to thank the anonymous referee for comments that helped to improve our paper.
TAAS gratefully acknowledges funding from the Natural Sciences and Engineering Research Council of Canada (NSERC) through the Discovery Grant program. 
\end{acknowledgments}

\newpage
\appendix

This Appendix summarizes the total disk masses and viewing inclinations derived in this work from \ha\ profile fitting for each of the 159 CAe stars reported by \citet{Anusha2021}. For each star, Table~\ref{tab:appendix} lists its LAMOST identification, its Gaia DR3 identification, its spectral type, its disk mass, $\log_{10}({M_{\rm disk}/M_*})$, expressed as a mean $\pm\,1$ standard deviation, and its viewing inclination (in degrees), expressed as a mean and $\pm \,1$ standard deviation. Viewing inclination standard deviations below $5^\circ$ were set to $5^\circ$, the spacing of the inclination grid used for the fits. Individual stars have comments to indicate detected \caII\ emission, cluster membership, or other information; see the table notes.

\startlongtable
\begin{deluxetable}{rrcrrl}
\tablewidth{0pt}
\tablecaption{Identifications and parameters for the 159 CAe stars.\label{tab:appendix}}
\tablehead{
\colhead{LAMOST ID} & \colhead{GAIA DR3} & \colhead{Spectral Type} & \colhead{$\log_{10}(M_{\rm disk}/M_*)$} & \colhead{Inclination($^\circ$)} & \colhead{Comment}
}
\startdata
 J001742.37+454454.4 &   386088972610039936 &  A3 & $ -7.27\pm0.46$ & $ 76\pm5.3$ & EB \\ 
 J004927.87+551549.0 &   417676983908240896 &  A0 & $ -9.51\pm0.33$ & $ 33\pm5.0$ &  \\ 
 J005911.26+555720.1 &   423539682985511808 &  A3 & $ -7.18\pm0.46$ & $ 80\pm5.0$ & EB \\ 
 J011345.53+383819.9 &   370088780617590912 &  A2 & $ -8.71\pm0.26$ & $ 43\pm5.0$ &  \\ 
 J011452.80+542305.3 &   411516523332160640 &  A0 & $-10.00\pm0.37$ & $ 39\pm5.0$ &  \\ 
 J011650.55+492723.3 &   403243797841227520 &  A0 & $ -9.97\pm0.20$ & $ 70\pm5.0$ &  \\ 
 J012208.07+231347.2 &   292928795495895936 &  A3 & $ -7.92\pm0.39$ & $ 66\pm5.0$ &  \\ 
 J020337.34+284455.1 &   299913516126375808 &  A2 & $ -8.39\pm0.31$ & $ 69\pm5.0$ &  \\ 
 J021331.68+561900.8 &   456990262666733312 &  A0 & $ -8.75\pm0.15$ & $ 79\pm5.0$ &  \\ 
 J021405.28+584915.9 &   506848991819829248 &  A0 & $ -8.85\pm0.24$ & $ 27\pm5.0$ &  \\ 
 J030757.57+262544.2 &   112445993319377536 &  A1 & $ -9.13\pm0.36$ & $ 40\pm9.3$ &  \\ 
 J031108.31+530719.3 &   446251057724879616 &  A0 & $ -9.58\pm0.40$ & $ 32\pm8.0$ &  \caII\ \\ 
 J032758.90+490110.3 &   441415165068758400 &  A1 & $ -8.80\pm0.25$ & $ 49\pm5.0$ &  \\ 
 J033617.64+585358.3 &   450016850691823744 &  A1 & $ -9.04\pm0.25$ & $ 45\pm5.0$ &  \\ 
 J034018.06+580620.4 &   449177098686670080 &  A2 & $ -8.54\pm0.36$ & $ 63\pm5.0$ &  \\ 
 J034056.31+554904.8 &   445653782391673856 &  A0 & $ -9.40\pm0.18$ & $ 78\pm5.0$ &  \\ 
 J034201.82+574616.0 &   449107348416724096 &  A0 & $-10.17\pm0.44$ & $ 28\pm5.0$ &  \\ 
 J034245.66+453622.3 &   244772630040359296 &  A1 & $ -9.78\pm0.31$ & $ 52\pm5.0$ &  \\ 
 J034349.11+530505.6 &   444034133044347776 &  A2 & $ -7.52\pm0.26$ & $ 53\pm5.0$ &  \\ 
 J034409.83+085740.2 &  3278563908174223744 &  A3 & $ -7.53\pm0.21$ & $ 70\pm5.0$ &  \caII\ \\ 
 J034409.84+085740.0 &  3278563908174223744 &  A2 & $ -8.60\pm0.22$ & $ 65\pm5.0$ &  \caII\ \\ 
 J034512.90+365346.7 &   223060878865985920 &  A4 & $ -7.44\pm0.44$ & $ 63\pm5.0$ & dS$^{a}$ \\ 
 J034638.80+494703.7 &   251076920477156480 &  A2 & $ -8.69\pm0.29$ & $ 69\pm5.0$ &  \\ 
 J035401.52+282155.3 &   166931295601204992 &  A2 & $ -8.43\pm0.11$ & $ 53\pm5.0$ &  \caII\ \\ 
 J035941.19+520908.3 &   251783833434584448 &  A4 & $ -7.55\pm0.41$ & $ 48\pm6.7$ &  Cl(UBC51)$^{c}$, dS$^{b}$ \\ 
 J040446.68+485013.9 &   247214988966460928 &  A1 & $-10.02\pm0.44$ & $ 19\pm6.5$ &  \\ 
 J040905.93+432827.6 &   232267884301477888 &  A1 & $-10.26\pm0.45$ & $ 20\pm5.0$ &  \\ 
 J041407.87+505417.6 &   271471349343947008 &  A1 & $-10.14\pm0.38$ & $ 22\pm5.0$ &  Cl(NGC1528)$^{d}$ \\ 
 J041606.86+412248.8 &   228712716527480960 &  A0 & $ -9.79\pm0.25$ & $ 67\pm5.0$ &  \\ 
 J041800.76+431127.1 &   229430624602079104 &  A0 & $-10.14\pm0.40$ & $ 27\pm5.0$ &  \\ 
 J041857.79+504651.2 &   271489972322885632 &  A0 & $-10.52\pm0.38$ & $ 23\pm5.0$ &  \\ 
 J042121.70+451929.7 &   232752837644721920 &  A0 & $ -9.49\pm0.17$ & $ 67\pm5.0$ &  \\ 
 J042242.03+463226.2 &   257070839396201984 &  A0 & $ -9.59\pm0.36$ & $ 65\pm5.0$ &  Cl(HSC1232)$^{e}$ \\ 
 J042447.61+464818.0 &   257095475328409856 &  A0 & $ -9.80\pm0.31$ & $ 47\pm5.0$ &  \\ 
 J043058.53+373855.6 &   178330585480057088 &  A1 & $ -8.09\pm0.16$ & $ 43\pm5.0$ &  \\ 
 J043603.65+422507.8 &   204289260440930176 &  A1 & $ -9.37\pm0.35$ & $ 39\pm5.5$ &  \\ 
 J044035.14+474907.3 &   257444501552180352 &  A2 & $ -9.81\pm0.29$ & $ 43\pm5.0$ &  \\ 
 J044257.98+431516.3 &   204671684328829696 &  A0 & $-10.31\pm0.39$ & $ 32\pm5.0$ &  \\ 
 J044350.95+430417.2 &   204483186804147456 &  A0 & $ -8.36\pm0.25$ & $ 48\pm5.0$ &  \\ 
 J044403.85+244323.2 &   147318619499009408 &  A0 & $ -9.19\pm0.28$ & $ 21\pm5.0$ &  \caII\ \\ 
 J044808.49+413832.8 &   203441438197027072 &  A0 & $-10.13\pm0.36$ & $ 22\pm5.0$ &  \\ 
 J044950.48+422202.6 &   203684945663121792 &  A0 & $ -8.39\pm0.24$ & $ 63\pm5.0$ &  \\ 
 J045210.76+545906.1 &   273529802612288384 &  A0 & $ -9.40\pm0.19$ & $ 27\pm5.0$ &  \\ 
 J045229.69+412503.7 &   200401387329324032 &  A0 & $ -9.28\pm0.16$ & $ 43\pm5.0$ &  \\ 
 J045623.87+341801.5 &   161764445645748096 &  A0 & $ -9.68\pm0.35$ & $ 29\pm5.0$ &  \\ 
 J045817.91+411510.8 &   201645454671050880 &  A2 & $ -8.65\pm0.19$ & $ 68\pm5.0$ &  \\ 
 J045916.91+401713.1 &   200783712434242432 &  A0 & $ -9.08\pm0.19$ & $ 76\pm5.0$ &  \\ 
 J050247.21+495552.7 &   256383678983504768 &  A0 & $ -9.10\pm0.26$ & $ 50\pm5.0$ &  \caII\ \\ 
 J050830.49+520025.8 &   265767976370499328 &  A1 & $ -8.83\pm0.33$ & $ 64\pm5.0$ &  \\ 
 J050837.62+474213.5 &   207141569700929792 &  A0 & $ -9.15\pm0.15$ & $ 63\pm5.0$ &  \\ 
 J051004.42+213405.0 &  3414968874039603200 &  A1 & $ -9.19\pm0.33$ & $ 28\pm5.0$ &  \\ 
 J051110.26+280651.4 &  3422096698684386304 &  A0 & $ -9.45\pm0.25$ & $ 58\pm5.0$ &  \\ 
 J051149.75+450007.9 &   205664612050187136 &  A0 & $-10.15\pm0.35$ & $ 34\pm5.0$ &  \\ 
 J051250.99+424524.1 &   201461625773564160 &  A2 & $ -7.36\pm0.40$ & $ 78\pm5.0$ &  \\ 
 J051256.73+472841.3 &   212902357795155456 &  A2 & $ -9.65\pm0.44$ & $ 15\pm6.1$ &  \\ 
 J051339.97+413555.0 &   201152632945311616 &  A1 & $ -8.88\pm0.30$ & $ 59\pm5.0$ &  \\ 
 J051359.86+330516.7 &   181854863845613824 &  A2 & $ -8.32\pm0.30$ & $ 60\pm5.0$ &  \\ 
 J051650.26+210129.7 &  3413997043197200000 &  A0 & $ -9.87\pm0.37$ & $ 46\pm6.6$ &  \\ 
 J051742.58+534619.4 &   266377964804108544 &  A0 & $ -9.96\pm0.34$ & $ 34\pm5.0$ &  \\ 
 J052352.30+441213.6 &   207635422220485120 &  A0 & $-10.16\pm0.35$ & $ 23\pm5.0$ &  \\ 
 J052420.40+474753.5 &   212529211036458752 &  A0 & $ -9.58\pm0.30$ & $ 60\pm6.0$ &  \caII\ \\ 
 J052544.34+240812.1 &  3416513997113757696 &  A0 & $-10.91\pm0.39$ & $ 13\pm5.0$ &  \\ 
 J052557.27+462913.5 &   209193086599072896 &  A0 & $ -9.62\pm0.37$ & $ 51\pm5.7$ &  Cl(NGC1883)$^{d}$ \\ 
 J052851.08+323236.1 &  3448940789744265984 &  A2 & $ -8.21\pm0.46$ & $ 76\pm5.6$ &  \\ 
 J052901.42+351630.1 &   183248769714056320 &  A0 & $-11.28\pm0.47$ & $ 14\pm8.3$ &  \\ 
 J053022.08+320220.7 &  3448710304620697856 &  A2 & $ -8.72\pm0.23$ & $ 66\pm5.0$ &  \\ 
 J053221.28+113353.5 &  3340653634735058816 &  A0 & $ -9.52\pm0.38$ & $ 46\pm6.6$ &  \\ 
 J053330.79+442700.6 &   195889197765289216 &  A0 & $ -9.96\pm0.28$ & $ 71\pm5.0$ &  \\ 
 J053509.41+190452.1 &  3400931615246607616 &  A1 & $ -8.19\pm0.28$ & $ 46\pm5.2$ &  \\ 
 J053820.19+315340.2 &  3447968482165561472 &  A2 & $ -9.63\pm0.42$ & $ 12\pm7.5$ &  \\ 
 J053832.80+125620.5 &  3341225037186241280 &  A0 & $ -7.30\pm0.39$ & $ 63\pm9.2$ &  \caII\ \\ 
 J053849.56+150251.8 &  3395757828920176768 &  A0 & $-10.20\pm0.31$ & $ 21\pm5.0$ &  \\ 
 J053925.93+312526.4 &  3447730884571427584 &  A1 & $ -8.94\pm0.18$ & $ 50\pm5.0$ &  \\ 
 J054156.03+251623.4 &  3428872198574732672 &  A3 & $ -9.57\pm0.46$ & $ 58\pm6.1$ &  \\ 
 J054247.63+161649.8 &  3396358471507663488 &  A0 & $ -9.14\pm0.30$ & $ 34\pm5.0$ &  \\ 
 J054434.62+215152.5 &  3403247221094375424 &  A0 & $ -9.29\pm0.18$ & $ 42\pm5.0$ &  \\ 
 J054514.38+294431.0 &  3444293639426677760 &  A2 & $ -7.33\pm0.44$ & $ 66\pm5.0$ &  \\ 
 J054619.00+293547.1 &  3444281785315294720 &  A0 & $ -9.65\pm0.35$ & $ 55\pm5.6$ &  \\ 
 J054627.18+413022.2 &   192686217314274048 &  A0 & $ -9.49\pm0.24$ & $ 56\pm5.0$ &  \\ 
 J054759.46+312203.2 &  3445018355029143040 &  A1 & $ -9.35\pm0.32$ & $ 44\pm5.0$ &  \caII\ \\ 
 J054826.95+251010.3 &  3428960846700152064 &  A0 & $ -8.80\pm0.27$ & $ 69\pm5.0$ &  Cl(Teutsch57)$^{e}$ \\ 
 J055329.89+303847.4 &  3444067861586996992 &  A2 & $ -8.66\pm0.37$ & $ 47\pm6.2$ &  \\ 
 J055427.65+383642.1 &  3457778939088450560 &  A2 & $ -8.47\pm0.30$ & $ 52\pm6.6$ &  \\ 
 J055430.75+295739.1 &  3443705224609706112 &  A2 & $-10.61\pm0.50$ & $  7\pm5.7$ &  \\ 
 J055549.11+401013.0 &  3458561864381418752 &  A0 & $-10.36\pm0.39$ & $ 15\pm5.4$ &  \\ 
 J055642.85-021100.4 &  3025557516593019648 &  A1 & $ -9.77\pm0.36$ & $ 26\pm5.0$ &  \\ 
 J055645.74+280415.1 &  3431466118305060096 &  A1 & $ -9.56\pm0.33$ & $ 38\pm5.0$ &  \\ 
 J055833.34+241358.2 &  3424951855504813184 &  A1 & $ -9.65\pm0.40$ & $ 24\pm5.0$ &  \\ 
 J055909.05+254251.4 &  3429622409102503680 &  A1 & $ -9.72\pm0.36$ & $ 21\pm5.0$ &  \\ 
 J055925.24+420010.1 &   192171650168942208 &  A1 & $ -9.71\pm0.40$ & $ 21\pm5.0$ &  \\ 
 J060056.78+264537.3 &  3429980433276744192 &  A1 & $ -9.67\pm0.40$ & $ 23\pm5.0$ &  \\ 
 J060230.66+221423.9 &  3423692193135773952 &  A1 & $ -9.68\pm0.40$ & $ 24\pm5.0$ &  \\ 
 J060306.07+315552.5 &  3450033910462371584 &  A2 & $ -8.71\pm0.22$ & $ 68\pm5.0$ &  \\ 
 J060427.73+422956.2 &   960857817034570880 &  A0 & $ -8.99\pm0.18$ & $ 73\pm5.0$ &  \caII\ \\ 
 J060654.75+081038.8 &  3328513518834841600 &  A0 & $ -9.73\pm0.32$ & $ 37\pm5.0$ &  \\ 
 J060718.51+323118.3 &  3450532057946624384 &  A0 & $ -9.14\pm0.29$ & $ 47\pm5.0$ &  \\ 
 J061511.42+155717.6 &  3345821957860995840 &  A0 & $ -9.01\pm0.35$ & $ 59\pm5.0$ &  \\ 
 J061903.73+273723.1 &  3433136379544342016 &  A2 & $ -9.55\pm0.31$ & $ 78\pm5.2$ &  \\ 
 J061907.58+115039.0 &  3331710451972427520 &  A2 & $ -9.62\pm0.46$ & $ 14\pm10.1$ &  \\ 
 J062158.25+103757.1 &  3330549367694319872 &  A0 & $ -9.24\pm0.24$ & $ 68\pm5.0$ &  \\ 
 J062323.24+491345.6 &   970595542983768448 &  A1 & $ -8.75\pm0.28$ & $ 67\pm5.0$ &  \\ 
 J062556.00+281433.0 &  3433583227941917568 &  A0 & $ -9.73\pm0.34$ & $ 34\pm5.0$ &  \\ 
 J062812.14+185018.6 &  3372128843006543872 &  A2 & $ -8.78\pm0.51$ & $ 80\pm5.0$ &  \\ 
 J062925.82+193445.9 &  3372309639651922432 &  A2 & $-10.61\pm0.38$ & $ 19\pm8.2$ &  EB, \caII\ \\ 
 J062937.28+170816.1 &  3369499704546103680 &  A1 & $ -9.52\pm0.39$ & $ 28\pm5.0$ &  \\ 
 J063242.65+303129.8 &  3435757787063121792 &  A1 & $ -9.35\pm0.37$ & $ 33\pm5.4$ &  \\ 
 J063259.00+182031.4 &  3371278503906101760 &  A1 & $ -9.62\pm0.40$ & $ 20\pm5.0$ &  \\ 
 J063322.20+585049.3 &  1004488985342311808 &  A3 & $ -7.71\pm0.17$ & $ 62\pm5.0$ &  \\ 
 J063548.02+543044.4 &   994503804853600256 &  A2 & $ -8.85\pm0.29$ & $ 62\pm5.0$ &  \\ 
 J063632.54+031727.9 &  3130168866789940736 &  A0 & $ -9.25\pm0.34$ & $ 53\pm5.7$ &  \\ 
 J063756.10+123409.5 &  3352304854518212096 &  A0 & $-11.46\pm0.32$ & $  8\pm5.2$ &  \\ 
 J063848.95+212539.7 &  3378911558438716160 &  A1 & $ -8.52\pm0.16$ & $ 63\pm5.0$ &  \\ 
 J063903.36+014446.1 &  3126705508243536000 &  A1 & $ -8.55\pm0.21$ & $ 58\pm5.0$ &  Cl(Collinder110)$^{d}$ \\ 
 J063922.76+190504.9 &  3371621074792159744 &  A0 & $-10.05\pm0.34$ & $  6\pm5.0$ &  \\ 
 J064125.00-005858.0 &  3107326654458643072 &  A0 & $-10.10\pm0.29$ & $ 30\pm5.1$ &  \\ 
 J064145.35+185543.0 &  3371558638850689408 &  A3 & $ -8.56\pm0.26$ & $ 70\pm5.0$ &  \\ 
 J064331.00+084841.4 &  3134437003430146816 &  A0 & $ -7.63\pm0.32$ & $ 54\pm5.0$ &  EB, \caII\ \\ 
 J064408.73+342007.4 &   938771140991006848 &  A2 & $ -9.01\pm0.34$ & $ 74\pm5.0$ &  \caII\ \\ 
 J064556.23+054333.0 &  3132372120593601536 &  A0 & $ -8.64\pm0.23$ & $ 68\pm5.0$ &  \caII\ \\ 
 J064627.17+015819.2 &  3125988729739626368 &  A3 & $ -9.11\pm0.49$ & $ 11\pm7.6$ &  Cl(Collinder115)$^{d}$ \\ 
 J064647.65+165123.4 &  3358352610846311808 &  A0 & $ -8.71\pm0.36$ & $ 51\pm5.4$ &  \caII\ \\ 
 J064714.86+035830.0 &  3127458501910175744 &  A0 & $ -9.39\pm0.29$ & $ 73\pm5.0$ &  \\ 
 J064729.66+104021.2 &  3350851334503166336 &  A0 & $ -7.79\pm0.14$ & $ 68\pm5.0$ &  \\ 
 J064752.48+043909.3 &  3129190340102731008 &  A1 & $ -8.84\pm0.20$ & $ 51\pm5.2$ &  \\ 
 J064912.02+030723.6 &  3126582225500375936 &  A3 & $ -7.75\pm0.38$ & $ 50\pm7.1$ &  \\ 
 J064931.23+124804.0 &  3351763959216420096 &  A0 & $ -9.51\pm0.35$ & $ 42\pm5.0$ &  \caII\ \\ 
 J065311.55+113256.3 &  3351386831024112896 &  A1 & $ -7.95\pm0.41$ & $ 76\pm5.2$ &  \\ 
 J065550.57+070949.9 &  3133003888803496960 &  A2 & $ -7.60\pm0.35$ & $ 79\pm5.0$ &  \\ 
 J065603.07+045352.3 &  3129443738871780992 &  A2 & $ -8.14\pm0.11$ & $ 65\pm5.0$ &  \\ 
 J065607.59+190904.3 &  3364737238648912512 &  A2 & $ -8.96\pm0.33$ & $ 68\pm5.0$ &  \\ 
 J065611.96+052217.5 &  3129520880784604928 &  A2 & $ -8.53\pm0.15$ & $ 62\pm5.7$ &  \\ 
 J065752.84+211422.2 &  3366338917850872960 &  A0 & $ -9.56\pm0.31$ & $ 51\pm5.0$ &  \\ 
 J065937.20+044253.6 &  3128648654531712640 &  A2 & $ -8.22\pm0.10$ & $ 59\pm5.0$ &  \\ 
 J070155.35+195449.8 &  3365202327769347712 &  A1 & $ -8.25\pm0.22$ & $ 56\pm5.0$ &  \\ 
 J070248.72+263829.4 &   883888330895443968 &  A0 & $ -9.17\pm0.27$ & $ 68\pm5.0$ &  \\ 
 J070253.50+582707.9 &  1002166194012384896 &  A0 & $ -9.05\pm0.22$ & $ 57\pm5.0$ &  \caII\ \\ 
 J070708.45+154039.7 &  3360013393098736128 &  A0 & $ -9.38\pm0.25$ & $ 68\pm5.0$ &  \caII\ \\ 
 J072256.82+231852.5 &   869391167003728768 &  A0 & $ -8.86\pm0.21$ & $ 68\pm5.0$ &  \\ 
 J072753.15+100219.1 &  3155598341654984192 &  A0 & $ -8.29\pm0.15$ & $ 59\pm5.0$ &  \caII\ \\ 
 J073534.71+365324.6 &   895645855405600640 &  A2 & $ -8.68\pm0.21$ & $ 52\pm5.7$ &  \\ 
 J074130.52+180030.2 &   671170759775402880 &  A0 & $ -9.27\pm0.30$ & $ 55\pm6.4$ &  \\ 
 J074244.51+353401.3 &   894886161592096640 &  A2 & $ -9.67\pm0.00$ & $ 75\pm5.0$ &  \caII\ \\ 
 J080138.51+054841.7 &  3095960830964919424 &  A0 & $ -8.73\pm0.39$ & $ 67\pm5.0$ &  \caII\ \\ 
 J080343.24-054731.7 &  3067380469312478464 &  A2 & $ -8.41\pm0.45$ & $ 80\pm5.0$ &  \\ 
 J081416.65+041330.4 &  3091666580925324416 &  A1 & $ -8.82\pm0.21$ & $ 65\pm5.0$ &  \\ 
 J081715.99-043855.1 &  3066218040704290816 &  A1 & $ -8.20\pm0.23$ & $ 45\pm5.0$ &  \\ 
 J090713.03+120019.8 &   604270738560210816 &  A0 & $ -9.06\pm0.40$ & $ 74\pm5.0$ &  \caII\ \\ 
 J100103.76+002434.3 &  3833574627431632128 &  A2 & $ -8.71\pm0.38$ & $ 68\pm5.0$ &  \\ 
 J112045.14+362535.6 &   760463232938062464 &  A4 & $ -7.57\pm0.48$ & $ 63\pm5.0$ & SB \\ 
 J114805.60+412843.2 &   768416142275956224 &  A1 & $ -8.36\pm0.28$ & $ 62\pm5.0$ &  \\ 
 J164317.98+072016.2 &  4442056358087613824 &  A1 & $ -7.89\pm0.19$ & $ 56\pm5.0$ &  \caII\ \\ 
 J173015.77+140534.4 &  4542547941297979136 &  A3 & $ -7.94\pm0.33$ & $ 60\pm5.0$ &  \\ 
 J173033.93+234132.3 &  4569535385482331904 &  A3 & $ -9.52\pm0.27$ & $ 61\pm5.2$ &  \\ 
 J173901.62+535717.0 &  1417669787018896384 &  A3 & $ -7.49\pm0.48$ & $ 73\pm6.6$ &  \\ 
 J184640.36+425427.9 &  2104833218301107072 &  A1 & $ -9.11\pm0.34$ & $ 52\pm5.8$ &  \\ 
 J194424.27+280911.8 &  2031268120808177152 &  A1 & $ -8.95\pm0.34$ & $ 46\pm5.0$ &  \\ 
 J200257.09+453007.9 &  2082192865140171520 &  A0 & $ -8.57\pm0.20$ & $ 62\pm5.0$ &  \caII\ \\ 
 J201118.79+263552.3 &  1836391783708353536 &  A1 & $ -8.68\pm0.31$ & $ 58\pm5.0$ &  \\ 
 J230200.43+523520.8 &  1989459676491228928 &  A0 & $-10.10\pm0.38$ & $ 23\pm5.0$ &  \caII\ \\ 
\enddata
\tablecomments{Spectral types adopted from \citet{Anusha2021}. Two stars (LAMOST IDs J034409.83+085740.0 and J034409.83+085740.2) are only reconcilable with a single Gaia DR3 ID and may represent the same star observed at two epochs separated by approximately 12 months. In the comments field, \caII\ indicates detected \caII\ IRT emission \citep{Anusha2021}; EB denotes eclipsing binaries \citep{Mowlavi2023}; dS denotes $\delta$ Scuti variables (\citealp[$^{a}$\!\!][]{Chen2020dS}, \citealp[$^{b}$\!\!][]{Guo2024dS}); SB denotes spectroscopic binaries \citep{GaiaMultip2023}; and Cl(ID) denotes probable cluster members as identified by $^{c}$\citet{Perren2023}, $^{d}$\citet{Cantat-Gaudin2018}, and $^{e}$\citet{Hunt2024}.}
\end{deluxetable}

\bibliography{maincitas}{}
\bibliographystyle{aasjournal}

\end{document}